\documentclass[aps,prl,preprint,superscriptaddress]{revtex4-2}

\usepackage{cancel}
\usepackage[version=4]{mhchem}
\usepackage{parskip}

\usepackage{graphicx}
\usepackage{multirow}
\usepackage{amsmath,amssymb,amsfonts}
\usepackage{amsthm}
\usepackage{mathrsfs}

\usepackage{hyperref}
\usepackage{scholax}
\usepackage[scaled=1.075,ncf,vvarbb]{newtxmath}
\usepackage{mathrsfs}
\usepackage{setspace}

\newcommand{\mrm}{\mathrm}

\newcommand{\ket}[1]{{\left| {#1} \right\rangle}}
\newcommand{\bra}[1]{{\left\langle {#1} \right|}}

\newcommand{\avg}[1]{\left\langle {#1} \right\rangle }

\newcommand{\Evec}{\vec{\mathcal{E}}}

\newcommand{\Esca}{\mathcal{E}}

\newcommand{\Bsca}{\mathcal{B}}

\newcommand{\eMax}{\ensuremath{e_{\text{max}}}}

\begin{document}
\title{Limit on the nuclear Schiff moment of $^{153}$Eu}

\author{Bassam Nima}
\affiliation{Department of Physics, University of Toronto, Toronto, ON M5S 1A7, Canada}

\author{Mingyu Fan}
\affiliation{Department of Physics, University of Toronto, Toronto, ON M5S 1A7, Canada}

\author{Xubo Wang}
\affiliation{Department of Chemistry, Johns Hopkins University, Baltimore, MD 21218, USA}

\author{Sen Wang}
\affiliation{Department of Chemistry, Johns Hopkins University, Baltimore, MD 21218, USA}

\author{En Fu Zhou}
\affiliation{School of Physics and Astronomy, Sun Yat-sen University, Zhuhai 519082, P.R. China}

\author{Andrew M. Jayich}
\affiliation{Department of Physics, University of California Santa Barbara, Santa Barbara, CA 93106, USA}

\author{Jiang Ming Yao}
\affiliation{School of Physics and Astronomy, Sun Yat-sen University, Zhuhai 519082, P.R. China}

\author{Lan Cheng}
\affiliation{Department of Chemistry, Johns Hopkins University, Baltimore, MD 21218, USA}

\author{Amar Vutha}
\affiliation{Department of Physics, University of Toronto, Toronto, ON M5S 1A7, Canada}

\begin{abstract}
The Schiff moment of a nucleus is a symmetry-violating nuclear moment that indicates new physics beyond the Standard Model. We place the limit, $|\mathscr{S}({}^{153}$Eu)$| < 1.7 \times 10^{-8}$ $e\,$fm$^3$ (95\% confidence), on the Schiff moment of the $^{153}$Eu nucleus, using nuclear spin resonances in two ensembles of oppositely-polarized $^{153}$Eu$^{3+}$ ions in a Y${}_2$SiO${}_5$ crystal. This measurement using octupolar nuclei in a mm-scale crystal constrains new physics at the TeV energy-scale.
\end{abstract}
\maketitle

The matter-antimatter imbalance of the Universe cannot be explained within the Standard Model of particle physics \cite{Morrissey_2012}. This motivates the search for new physics that breaks time-reversal (T) symmetry \cite{Sakharov_1967, khriplovich_2012}. New sources of T-violation that couple to quarks and gluons in nuclei induce symmetry-violating nuclear moments, whose existence can be detected by precision measurements of nuclear spin transitions \cite{Wilkening1984,Graner2016,Bishof2016,Allmendinger2019,Zheng2022}. One such quantity is the nuclear Schiff moment (NSM), an electrostatic moment that is odd under parity (P) and T, and which can be generated by T-violation in nuclei with spin $I \geq 1/2$ \cite{Schiff:1963,Engel2025,Vutha2026}. Larger NSMs are induced in octupolar nuclei than in spherical nuclei \cite{Auerbach1996,Spevak1997,EngelFriarHayes2000}. 

Our measurements use $^{153}$Eu ($I=5/2$), chosen for its octupolar enhancement of T-violation effects. The NSM of $^{153}$Eu interacts with the electrons in electrically-polarized Eu$^{3+}$ ions producing P,T-odd hyperfine energy shifts. The measurements reported here were performed with Eu$^{3+}$ ions that are statically polarized when doped into non-centrosymmetric sites in the Y$_2$SiO$_5$ (YSO) crystal.

Four distinct Eu${}^{3+}$ dopant positions, each coordinated to seven O${}^{2-}$ ions (``site 1''), occur within the primitive cell of YSO. Site 1 ions are selected using their 580.04 nm optical transition between the electronic ground ${}^7$F${}_0$ and excited ${}^5$D${}_0$ states. Two crystal symmetry operations relate the four dopant positions, an inversion $\hat{P}: (x,y,z) \rightarrow (-x,-y,-z)$ with eigenvalues $\rho=\pm1$ and a reflection $\hat{\Sigma}: (x,y,z) \rightarrow (-x,-y,z)$ with eigenvalues $\sigma=\pm1$. The effective Hamiltonian for the nuclear spin degree of freedom in the $^7F_0$ electronic state of ${}^{153}$Eu${}^{3+}$ ions depends on the dopant position through these two crystal symmetries \cite{Nima2025}. In $\hbar=1$ units, the effective Hamiltonian is
\begin{equation}\label{eq:hamiltonian}
\begin{split}
    H(\sigma, \rho) = \sum_{i,j} Q_{ij} I_i I_j - \sigma \, (\mu_x \Bsca_x + \mu_y \Bsca_y) - \mu_z \Bsca_z - \sigma \rho D \hat{n} \cdot \vec{\mathcal{E}} \, \mathbb{I} + \rho \, \Omega \, \hat{I} \cdot \hat{n}.
\end{split}
\end{equation}
The spatial indices $i, j \in (x,y,z)$ where $x,y,z$ are aligned to the $D_1, D_2, b$ dielectric axes of the crystal. $Q_{ij}$ is the electric quadrupole interaction tensor, proportional to the product of the nuclear quadrupole moment and the electric field gradient at the nucleus. $\mu_i = -\sum_{j} M_{ij} I_j$ is the magnetic moment, $I_i$ is the nuclear spin, $\mathcal{B}_i$ and $\mathcal{E}_i$ are the lab magnetic and electric fields respectively, $\rho \sigma D \hat{n}$ is the electric dipole moment of the $^{153}$Eu$^{3+}$ ion in the crystal that is polarized along the unit vector $\hat{n}$. The $\rho \, \Omega\hat{I}\cdot\hat{n}$ term is the P,T-odd hyperfine interaction. The parameter $\Omega = \mathscr{S} \mathcal{W}_\mrm{NSM}$ involves the NSM, $\vec{\mathscr{S}} = \mathscr{S} \hat{I}$, and a crystal-aligned vector that is proportional to the electron density gradient at the nucleus of the  $^{153}$Eu$^{3+}$ ion, $\vec{\mathcal{W}}_\mrm{NSM} = \rho \mathcal{W}_\mrm{NSM} \hat{n}$. The values of $\mathscr{S}$ and $\vec{\mathcal{W}}_\mrm{NSM}$ are obtained from nuclear and molecular calculations respectively.

\subsection*{Calculation of $\mathscr{S}$} \label{sec:nuclear}
We performed multi-reference covariant density functional theory (MR-CDFT) calculations to evaluate the NSM of $^{153}$Eu \cite{methods}, using a similar method as in Ref.\ \cite{Zhou:2025_short}. 
We first carried out a self-consistent CDFT calculation to generate a set of mean-field configurations with different quadrupole and octupole deformation parameters, $(\beta_2,\beta_3)$, for $^{153}$Eu with the false quasiparticle vacuum approximation~\cite{Zhou:2024PRC}. We then used particle-number, parity and angular momentum projection to restore the missing quantum numbers (nucleon numbers $N, Z$ and spin-parity $J^\pi$) of the symmetry-breaking states. We found the ground state $J^\pi=5/2^+$ to have $(\beta_2,\beta_3)\approx(0.40,0.08)$, indicating that it has a prolate quadrupole deformation with appreciable octupole deformation. The excited state $J^\pi=5/2^-$ (97.4 keV) was found to have
$(\beta_2,\beta_3)\approx(0.39,0.075)$.

The wavefunctions of nuclear states were constructed by mixing the wavefunctions of intrinsic states on the $(\beta_2,\beta_3)$ deformation plane, with the mixing weights determined variationally. These wavefunctions incorporate beyond-mean-field correlations associated with symmetry restoration, as well as both static and dynamical quadrupole--octupole deformations. Using these wavefunctions, energies, magnetic moments, quadrupole moments and transition matrix elements were calculated and benchmarked against experimental values where available \cite{methods} -- the calculated values showed good agreement  with published values for nuclear moments and transition matrix elements.

The nuclear Schiff moment of the nuclear ground state induced by a P,T-violating pion-nucleon interaction $V_{PT}$ (which in turn arises from physics beyond the Standard Model) is \cite{Engel:2013PPNP}
\begin{eqnarray}
\label{eq:Schiff_moment_def}
\mathscr{S} &=& \sum_{k\ne 0} \frac{\bra{ \Psi_0} {\mathscr{S}}_z \ket{\Psi_k} \bra{\Psi_k} V_{PT} \ket{\Psi_0}}{E_0-E_k} + c.c. \equiv g_{\pi NN} \sum_{\alpha=0}^2 \bar g_{\pi}^{(\alpha)} a_\alpha,
\end{eqnarray}
where the structure factors $a_\alpha$ are defined by the last relation. The QCD pion-nucleon coupling strength is $g_{\pi N N }\approx 12.9$, and the isoscalar ($\bar{g}_{\pi}^{(0)}$), isovector ($\bar{g}_{\pi}^{(1)}$), and isotensor ($\bar{g}_{\pi}^{(2)}$) coupling parameters arise from P,T-violating new physics. $\ket{\Psi_k}$ and $E_k$ denote the wavefunction and energy of the $k$-th nuclear excited state, respectively. $\mathscr{S}_z$ is the component of the Schiff moment vector along the nuclear spin axis. The resulting relation between the Schiff moment of $^{153}$Eu and the P,T-odd isospin couplings is
\begin{equation}\label{eq:schiff_calc}
\mathscr{S} = g_{\pi NN} \left( 0.96\, \bar{g}_{\pi}^{(0)} + 2.90\, \bar{g}_{\pi}^{(1)} + 1.87\, \bar{g}_{\pi}^{(2)} \right) \; e \,\mrm{fm}^3.
\end{equation}
This is the first beyond-mean-field microscopic calculation of the Schiff moment in a stable nucleus that has octupole-enhanced sensitivity to T-violating new physics \cite{EngelFriarHayes2000}. The semi-empirical calculations for the dependence of $\mathscr{S}$($^{153}$Eu) on $g_\pi^{(0)}$ in Refs.\ \cite{Sushkov2024, Flambaum2025}, based on a rigid quadrupole-octupole-deformed nuclear shape with $\beta_2=0.249, \beta_3=0.095$, are in rough agreement with our microscopic calculation.

\subsection*{Calculation of $\vec{\mathcal{W}}_\mrm{NSM}$}\label{sec:wnsm}
To compute $\vec{\mathcal{W}}_\mrm{NSM}$ in a doped crystal, while circumventing the uncertainty in the knowledge of the local structure of the dopant site, our calculations exploit the correlation between $\vec{\mathcal{W}}_\mrm{NSM}$ and the transition dipole moment of the ${}^7F_0-{}^5D_0$ optical transition, $\vec{d}_{00}$. Both quantities are sensitive to the strength and direction of the parity-odd crystal field acting on Eu$^{3+}$. In Eu:YSO, $\vec{d}_{00}$ has been measured to be strongly aligned with the $x$ axis \cite{konz_temperature_2003}. Therefore, the dominant component is expected to be $\vec{\mathcal{W}}_\mrm{NSM}\cdot \hat{x} = \mathcal{W}_\mrm{NSM} \cos\theta$, where $\theta$ is the angle between $\vec{\mathcal{W}}_\mrm{NSM}$ and $\vec{d}_{00}$.

We computed $\vec{\mathcal{W}}_\mrm{NSM}$ and $\vec{d}_{00}$ for a range of model systems using all-electron relativistic multireference techniques \cite{Wang2026} and an extensive basis set for Eu constructed specifically for accurate calculations of $\vec{\mathcal{W}}_\mrm{NSM}$ \cite{Chen24}. The set of 61 model systems studied here include the isolated Eu$^{3+}$ ion in uniform external electric fields, the EuO$^+$, Eu(OH)$^{2+}$, and Eu(OH)$_2^+$ molecular cations, the Eu(OH)$_3$ neutral molecule, and the Eu(OH)$_7^{4-}$ anion \cite{methods}. The calculations confirm a linear relationship $\mathcal{W}_\mrm{NSM} = \eta d_{00}$ in these species over a wide range of chemical environments, with $\eta = 6.4\times 10^6$ atomic units (a.u.) \cite{methods}. 

To estimate $\mathcal{W}_\mrm{NSM}$ in Eu:YSO, we use the results for the closest (and largest) model system, Eu(OH)$_7^{4-}$, where we computed $d_{00} = 7\times 10^{-4}$ a.u. (close to the value $d_{00} = 8 \times 10^{-4}$ a.u. measured in Eu:YSO \cite{Chen2025}), $\mathcal{W}_\mrm{NSM} = 3842$ a.u. and $\cos \theta=0.70$. Scaling the computed $\mathcal{W}_\mrm{NSM}$ in Eu(OH)$_7^{4-}$ by the ratio of the $d_{00}$ values for Eu:YSO and Eu(OH)$_7^{4-}$, we determine $\vec{\mathcal{W}}_\mrm{NSM}\cdot \hat{x}=3027 \; \mrm{a.u.} = 2\pi\times 134 ~\text{kHz}/(e ~\text{fm}^3)$ in Eu:YSO. We estimate an uncertainty of 20\% in this calculated value \cite{methods}.

\subsection*{Experiment scheme}
When T-symmetry is unbroken, the nuclear spin eigenstates in $^{153}$Eu:YSO occur as three degenerate Kramers-conjugate pairs, denoted as $a,\bar{a}, b,\bar{b}$ and $c,\bar{c}$. Due to magnetic fields and the NSM interaction, the Kramers degeneracy is lifted. The resonance frequency for the nuclear spin transition between the Kramers-conjugate states $\ket{m}$ and $\ket{\bar{m}}$ is
\begin{equation}
    f_0(\rho) = f_\mathrm{Z} + \rho \frac{2\Omega}{2\pi} \, \bra{m}\hat{I}\cdot\hat{n}\ket{m},
\end{equation}
where $f_Z$ is the Zeeman shift from the magnetic dipole interaction and $\rho = \pm 1$ for the two oppositely-polarized sub-ensembles of Eu$^{3+}$ ions. The NSM-induced hyperfine interaction parameter $\Omega$ can be extracted from the difference of the $\rho = \pm1$ resonance frequencies. Therefore the $\rho = \pm1$ sub-ensembles of Eu$^{3+}$ ions act as mutual comagnetometers, with identical Zeeman shifts and opposite NSM-induced shifts \cite{Nima2025}. We define the difference 
\begin{equation}\label{eq:fd}
    f_d = f_0(\rho=+1) - f_0(\rho=-1) = \frac{4\Omega}{2\pi} \, \bra{m}\hat{I}\cdot\hat{n}\ket{m}
\end{equation}
which is only sensitive to the P,T-odd parameter $\Omega$.

The $b,\bar{b}$ nuclear spin states used in this measurement have $\bra{b}\hat{I} \cdot \hat{x} \ket{b} = - \bra{\bar{b}}\hat{I} \cdot \hat{x} \ket{\bar{b}} = 0.254$.

\subsection*{Measurement of $\Omega$} 
We measured the resonance frequencies $f_0(\rho)$ for the $b-\bar{b}$  transition, which occurs at $f_0(\rho)\approx 230.5$ kHz in the static magnetic field $\Bsca_x \approx 350$ G applied to the Eu:YSO crystal. Each  measurement cycle consisted of a state-preparation step, radio-frequency (rf) spectroscopy step, and state-detection step. State-preparation using optical pumping, and state-detection using frequency-modulated optical absorption spectroscopy were performed using a laser tuned to the ${}^7F_0 - {}^5D_0$ optical transition \cite{methods}. The frequency of the laser, $\nu_L$, was used to select a spectral class ($\sim$ 1 MHz linewidth) out of the inhomogeneously-broadened ${}^7F_0 - {}^5D_0$ line ($\sim$ 1 GHz linewidth). 

A static electric field $\Evec_\mrm{DC} = \pm 33 \, \hat{x}$ V/cm was applied across the crystal in order to resolve the $\rho = \pm 1$ optical lines during the state-detection step. The sign of $\Esca_\mrm{DC}$ was reversed between consecutive measurements in order to cancel the effect of detuning-dependent optical power during state-detection \cite{methods}. Note that $\Esca_\mrm{DC}$ is only required during the state-detection step, in order to distinguish the nuclear spin state populations for the $\rho = \pm1$ sub-ensembles. The sign of $\Esca_\mrm{DC}$ exchanges the frequencies of the optical absorption lines corresponding to the $\rho = \pm 1$ sub-ensembles. The average of sequential $f_d$ measurements with $\pm \Esca_\mrm{DC}$, denoted as $\bar{f}_d$, is the principal quantity measured in the experiment. This quantity $\bar{f}_d$ has the same sensitivity to $\Omega$ as in Equation (\ref{eq:fd}) but has reduced sensitivity to fluctuations in the frequencies of the optical absorption lines. 

\begin{figure}
    \centering
    \includegraphics[width=\linewidth]{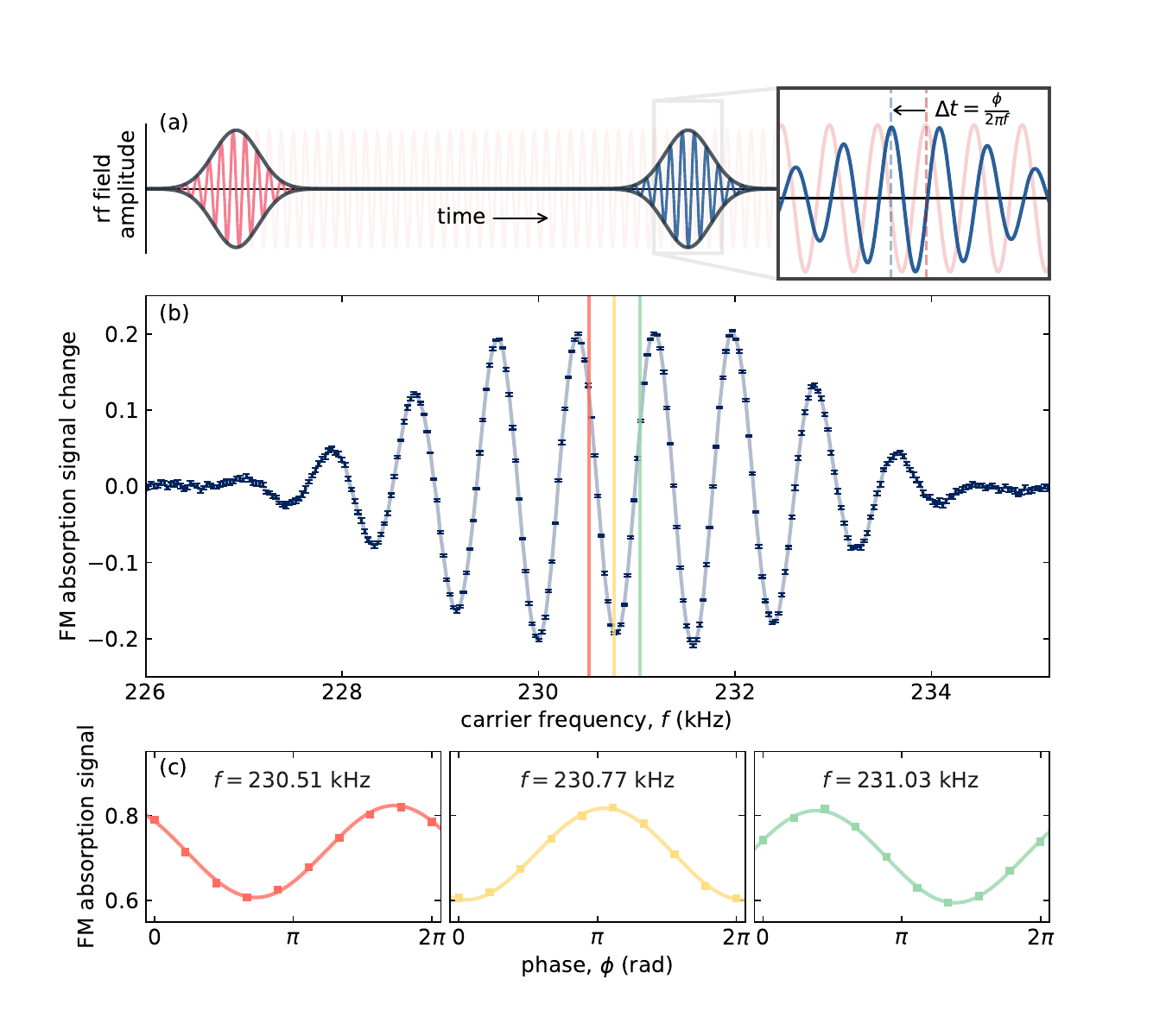}
    \caption{\textbf{Precision rf spectroscopy.} (a) Schematic timing diagram of the phasor Ramsey sequence, showing two pulses with a relative phase difference $\phi$. The carrier frequency of the pulses is shown in pale red for illustration.
    (b) Spectrum of the $b-\bar{b}$ nuclear spin transition. The traditional Ramsey resonance lineshape shows the population transferred between the nuclear spin states as a function of the carrier frequency of the rf pulses.
    (c) Phasor Ramsey method. Fixing the carrier frequency, the population transferred between nuclear spin states is measured versus $\phi$, and the phase offset of this sinusoidal signal is used to extract the resonance frequency. This method enables measurements of the resonance frequency with accuracies that are a small fraction of the resonance linewidth.}
    \label{fig:ramsey}
\end{figure}

The rf spectroscopy step was performed using the phasor Ramsey method \cite{methods,Vutha2015,Kato2018,Bezginov2019,Heydarizadmotlagh2024}, illustrated in Fig.\ \ref{fig:ramsey}. As in the traditional Ramsey method, two time-separated rf pulses with a variable phase difference $\phi$ were applied to the crystal to drive the $b-\bar{b}$ transition. The  $b-\bar{b}$ population transfer was measured for a range of different $\phi$ values, and the phase \textit{offset} of the sinusoidal curve of population transfer versus $\phi$ was used to extract the resonance frequency. 

\begin{figure}
    \centering
    \includegraphics[width=1\linewidth]{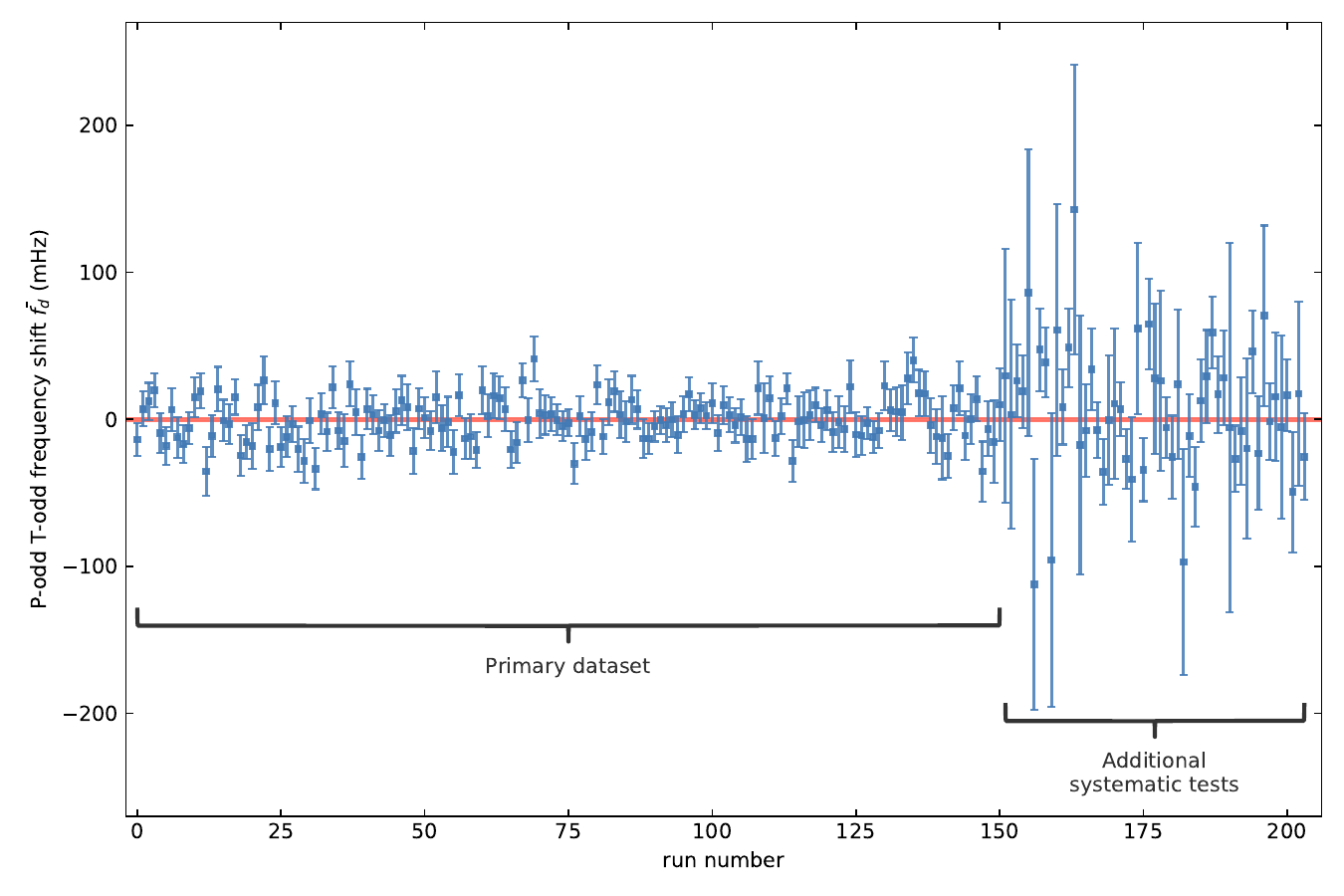}
    \caption{\textbf{Experimental dataset.} Data points are measurements of the P,T-odd linear combination of nuclear spin resonance frequencies, $\bar{f}_d$. The first section, up to run \# 150, shows the primary dataset acquired with the optimal set of parameters for minimizing statistical uncertainty. The second section is the dataset acquired with various (often sub-optimal) settings for studying  systematic errors. The solid line is a fit to a constant that yields $\bar{f}_d = -0.27 \pm 1.08_\mrm{stat}$ mHz.}
    \label{fig:f_d_bar}
\end{figure}

For the first part of the dataset shown in Fig.\ \ref{fig:f_d_bar}, we measured $\bar{f}_d$ for 562 hours with experiment settings optimized to yield the best precision. The time between rf Ramsey pulses, the optical power used for state-preparation and state-detection, and the laser frequency $\nu_L$ were periodically varied during these measurements. In the second portion of the dataset we measured $\bar{f}_d$ for 198 hours and introduced deliberate variations of experiment settings, even ones that led to decreased precision, in order to characterize systematic errors. 

There is no precedent for Schiff moment search using optical and rf spectroscopy in a solid-state system. Therefore, using this experiment as a pathfinder, we conducted studies of systematic errors by searching for possible correlations of $\bar{f}_d$ against a large number of passively-monitored as well as actively-controlled experimental parameters. These parameters included the rf carrier frequency ($f$) and the time interval between rf pulses in the phasor Ramsey measurement ($\tau_R$), the optical power used for state-preparation ($P_\mrm{prep}$) and absorption measurements ($P_\mrm{det}$), the optical detuning $\Delta \nu_L$, and the lab electric and magnetic fields. Consistency of the $\bar{f}_d$ measurements for different values of $f$ tests the lineshape-independence of the phasor Ramsey measurement, whereas consistency for different values of $\tau_R$ precludes systematic errors due to phase shifts or timing errors between the rf pulses. Varying the optical power parameters $P_\mrm{prep}$ and $P_\mrm{det}$ tests possible systematic effects from the state-preparation sequence and the detection process, and also verifies the independence of the rf spectroscopy step from the optical preparation/detection steps. The detuning $\Delta \nu_L$ selects the homogeneous spectral class of the ${}^7F_0-{}^5D_0$ transition -- and therefore the subset of Eu$^{3+}$ environments in the crystal -- that participates in the measurement. Thus, consistency of $\bar{f}_d$ for different laser detunings confirms the absence of any detectable systematic effects due to inhomogeneities in the crystal. The lab electric field changes the separation between the $\rho = \pm1$ optical absorption lines, and the lab magnetic field tunes the resonance frequency of the $b-\bar{b}$ transition: consistency of $\bar{f}_d$ against changes in these parameters indicates there are no significant systematics due to background electromagnetic fields.

All the measurements of $\bar{f}_d$ were masked by a computer-generated random number, $f_\mrm{mask}$, generated at the start of the experiment in March 2025 and stored in a machine-readable file. The value of $f_\mrm{mask}$ was chosen from a uniform distribution over the interval $[-100,+100]$ mHz, a hundredfold larger range than the precision of the measurement. This frequency mask was automatically added to the measured value of $f_0(\rho=+1)$ in the analysis code. Only the masked value of $\bar{f}_d$ was available to the experimenters until the conclusion of the experiment and the completion of the systematic error analysis summarized in Table \ref{tab:systematics}.

\begin{table}[h]
\centering
\caption{\textbf{Uncertainties and corrections to $\bar{f}_d$.} Note that the uncertainties from the two optical spectrum effects are added linearly to account for possible correlations, while other effects are added in quadrature, to obtain the total systematic uncertainty \cite{methods}.} \label{tab:systematics}
\begin{tabular}{lcc}
    ~\\
    \hline
    Effect & Correction (mHz) & Uncertainty (mHz) \\
    \hline
    Optical spectrum  &  & \\
    \hspace{6mm} laser frequency drifts & $-0.39$ & $0.30$   \\
    \hspace{6mm} electric field offset drifts & $-0.05$ & $0.14$ \\
    Crystal temperature  & $+0.01$ & $0.06$\\
    Laser power  & $-0.01$ & $0.02$\\
     Magnetic field  & & \\
    \hspace{6mm} $x$ & $0$ & $0.001$  \\
    \hspace{6mm} $y$ & $0$ & $0.003$ \\
    \hspace{6mm} $z$ & $0$ & $0.003$ \\
    AOM temperature  & $0$ & $0.001$ \\
    \hline
    \hline
    Total systematic & $-0.44$ & $0.44$ \\
    \hline
\end{tabular}
\end{table}

\begin{figure}[h!]
    \includegraphics[height=0.7\textheight]{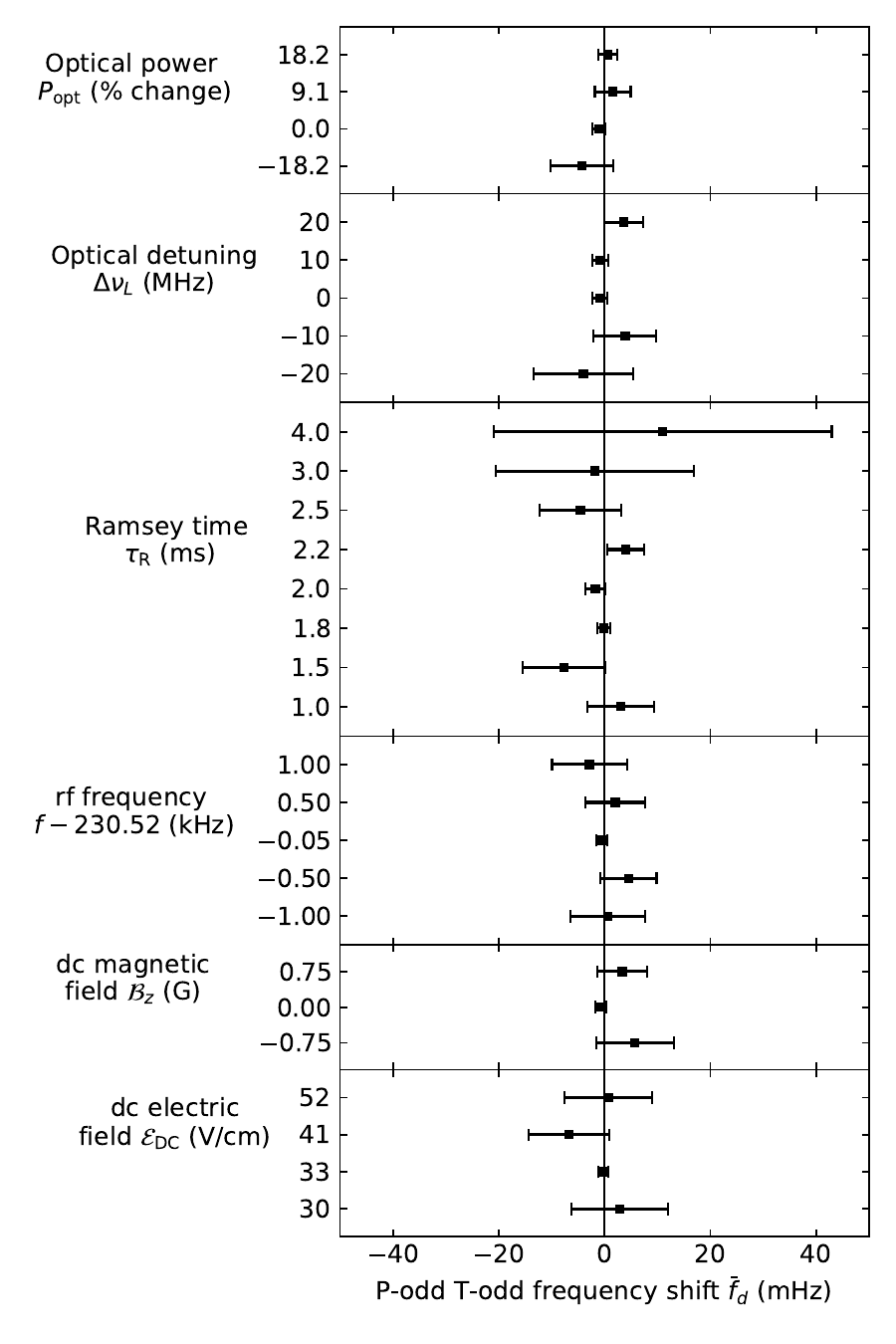}
    \caption{\textbf{Consistency of $\bar{f}_d$ measurements.} The data points show measurements of $\bar{f}_d$ acquired with various settings of the experimental parameters. Measurements with the smallest error bars correspond to the settings used in the primary dataset. }
    \label{fig:systematics}
\end{figure}

The dominant error in Table \ref{tab:systematics} is from a second-order technical source: electric field fluctuations (produced by the output voltage drift of an amplifier) and residual laser frequency fluctuations interacted with an optical-detuning-dependent background to the absorption measurements (produced by an acousto-optic modulator used for state-preparation and detection), which led to an apparent dependence of $\bar{f}_d$ on the frequencies of the $\rho = \pm1$ optical absorption lines. We applied a correction of $-0.44$ mHz for this effect (see \cite{methods} for a discussion of this and other systematic effects). The corrected values of $\bar{f}_d$ were consistent across variations of a large number of experimental parameters (Fig.\ \ref{fig:systematics}). The systematic error analysis from this pathfinder experiment indicates no fundamental barriers to a $\mu$Hz-level measurement of P,T-odd frequency shifts in a higher-precision experiment with an improved apparatus.

After the systematic error analysis was completed, the value of $f_\mrm{mask}$ was unmasked and the final result was found to be $\bar{f}_d = -0.27\pm1.08_\mathrm{stat}\pm 0.44_\mathrm{syst}$ mHz. The total uncertainty is less than 5 parts per million of the linewidth of the nuclear spin resonance (Fig.\ \ref{fig:ramsey}). The bound inferred from this measurement is $|\bar{f}_d| < 2.3$ mHz (95\% confidence \cite{Feldman1988}).

\subsection*{Implications of the measurement}
The resulting bound on the P,T-odd hyperfine parameter is $|\Omega| < 14.5$ mrad/s. Using the calculated value of $\mathcal{W}_\mrm{NSM}$, the  Schiff moment of $^{153}$Eu is $|\mathscr{S}| < 1.7 \times  10^{-8}$ $e\,$fm$^3$. 

\begin{figure}
    \centering
    \includegraphics[width=0.85\linewidth]{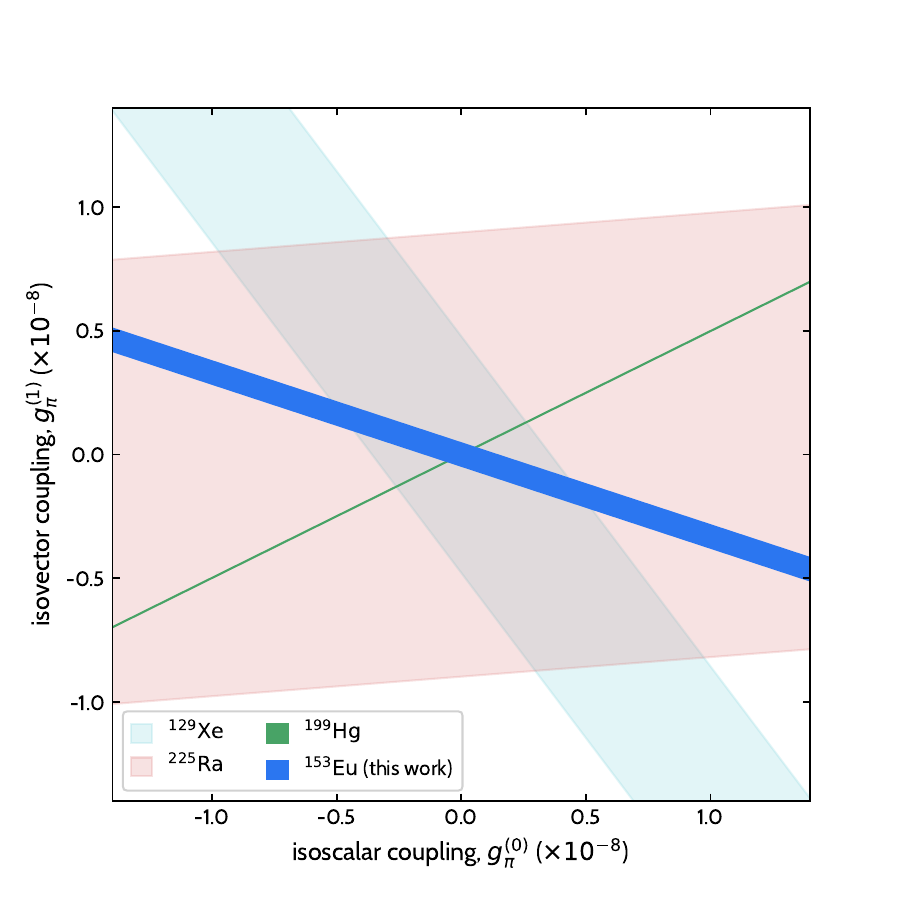}
    \caption{\textbf{Constraint on T-violating pion-nucleon couplings.} The coloured regions show the regions to which the isoscalar ($g_\pi^{(0)}$) and isovector ($g_\pi^{(1)}$) coupling constants are limited by experiments. 
This plot is a cut through the $g_\pi^{(2)}=0$ plane, for the purpose of illustration. In addition to the constraint from the present work, bounds derived from $^{199}$Hg \cite{Graner2016}, $^{225}$Ra \cite{Bishof2016} and $^{129}$Xe \cite{Allmendinger2019} are shown, using structure factors $a_i$ calculated for these nuclei in Ref.\ \cite{Zhou:2025_short}.}
    \label{fig:g0-g1-exclusion}
\end{figure}

This bound can be used to constrain a linear combination of the T-violating isospin couplings, $\bar{g}_\pi^{(i)}$, as shown in Fig.\ \ref{fig:g0-g1-exclusion}. Making he simplifying assumption that only the isoscalar $\bar{g}_\pi^{(0)}$ contributes to $\mathscr{S}$ leads to $| \bar{g}_\pi^{(0)} | < 1.3 \times 10^{-9}$. Using the relation $\bar{g}_\pi^{(0)} \approx -17.2 \times 10^{-3} \,  \bar{\theta}$ between $\bar{g}_\pi^{(0)}$ and the QCD $\bar{\theta}$-parameter \cite{Mulder2025} leads to the limit $|\bar{\theta}| \lesssim 8.2 \times 10^{-8}$. If we assume that only $\bar{g}_\pi^{(1)}$ contributes to $\mathscr{S}$, then $| \bar{g}_\pi^{(1)} | \lesssim 4.6 \times 10^{-10}$, which constrains the energy scale of T-violating new physics to be $M_{\cancel{T}} \gtrsim$ 9.7 TeV \cite{methods}.

In summary, we have measured the Schiff moment of a stable nucleus with octupolar enhancement, $^{153}$Eu, using precision spectroscopy of ions in a crystal. This measurement demonstrates a way to make significant inroads into the parameter space of new physics theories using octupolar nuclei in crystals.

\section*{Acknowledgments}
We acknowledge helpful discussions with Yoshiro Takahashi, David DeMille, Joseph Thywissen, Aephraim Steinberg, Eric Cornell and Daniel Comparat. The experiment was made possible by funding from the Alfred P. Sloan Foundation (Grant No. G-2023-21045) and the John Templeton Foundation (Grant No. 63119) through the Small-scale Experiments for Fundamental Physics program, and funding from an NSERC Discovery Grant (No. SAPIN-2021-00025). The work of E.F.Z. and J.M.Y. is supported in part by the National Natural Science Foundation of China (Grant Nos. 12405143, 12375119). L.C. acknowledges funding from the National Science Foundation (Grant No. PHY-230925).

\bibliography{eu_schiff}

\newpage

\renewcommand{\thefigure}{A\arabic{figure}}
\renewcommand{\thetable}{A\arabic{table}}
\renewcommand{\theequation}{A\arabic{equation}}
\renewcommand{\thepage}{A\arabic{page}}
\setcounter{figure}{0}
\setcounter{table}{0}
\setcounter{equation}{0}
\setcounter{page}{1} 

\section*{Appendix}

\section*{A1. Calculation of $\mathscr{S}$}
\begin{figure}[bt]
 \centering
\includegraphics[width=0.3\columnwidth]{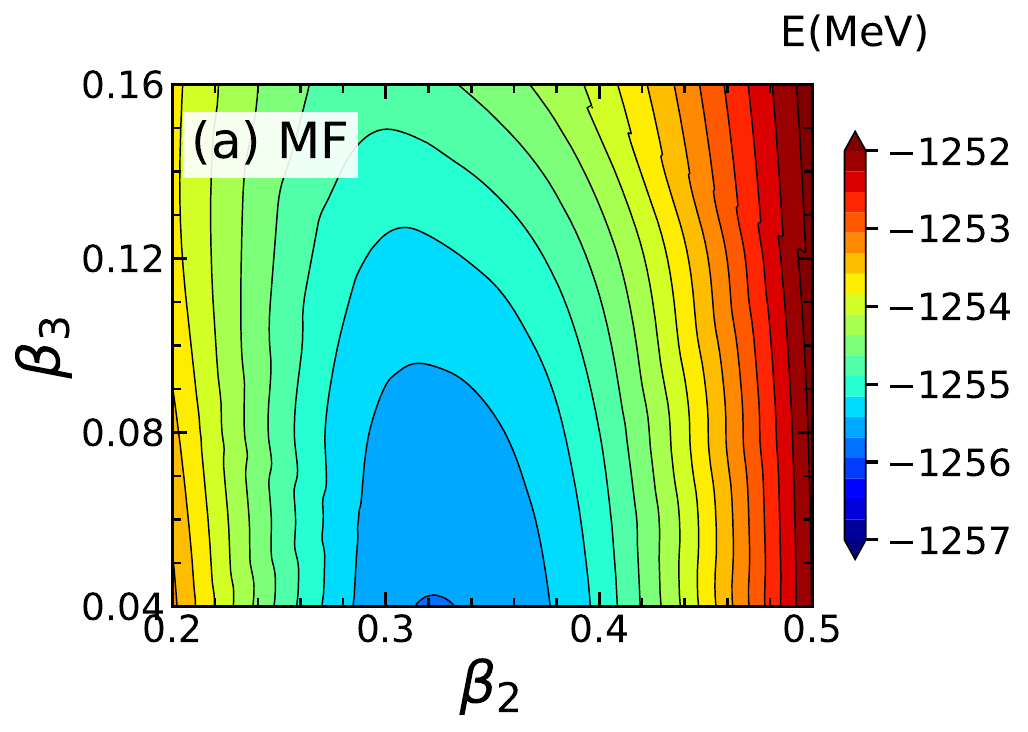}
\includegraphics[width=0.3\columnwidth]{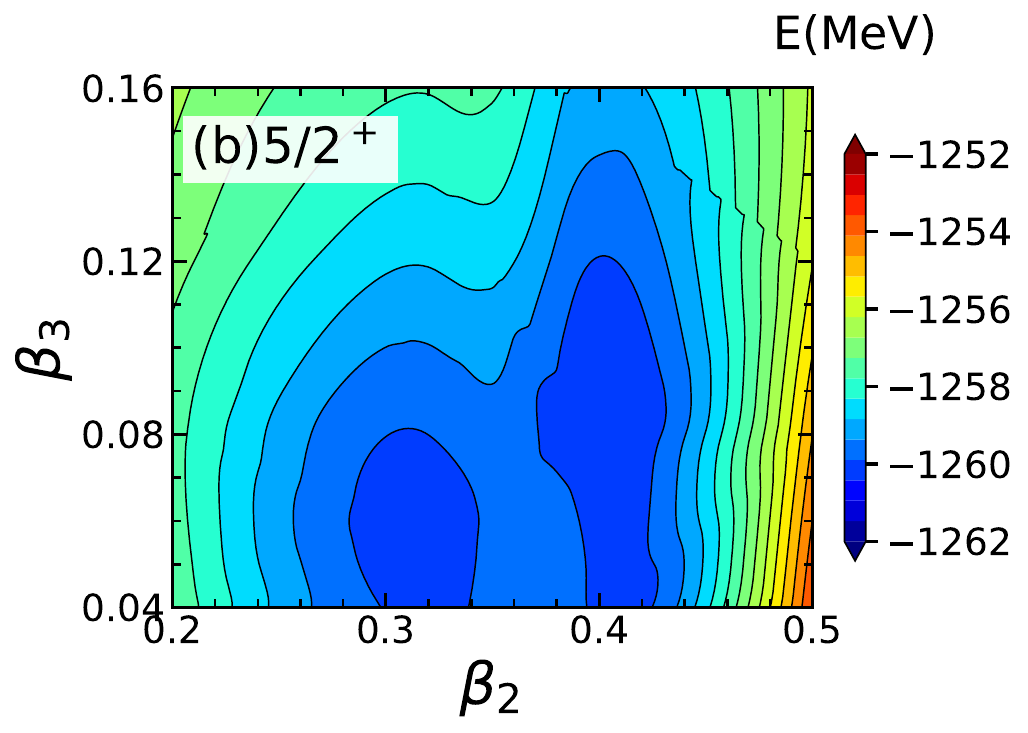}
\includegraphics[width=0.3\columnwidth]{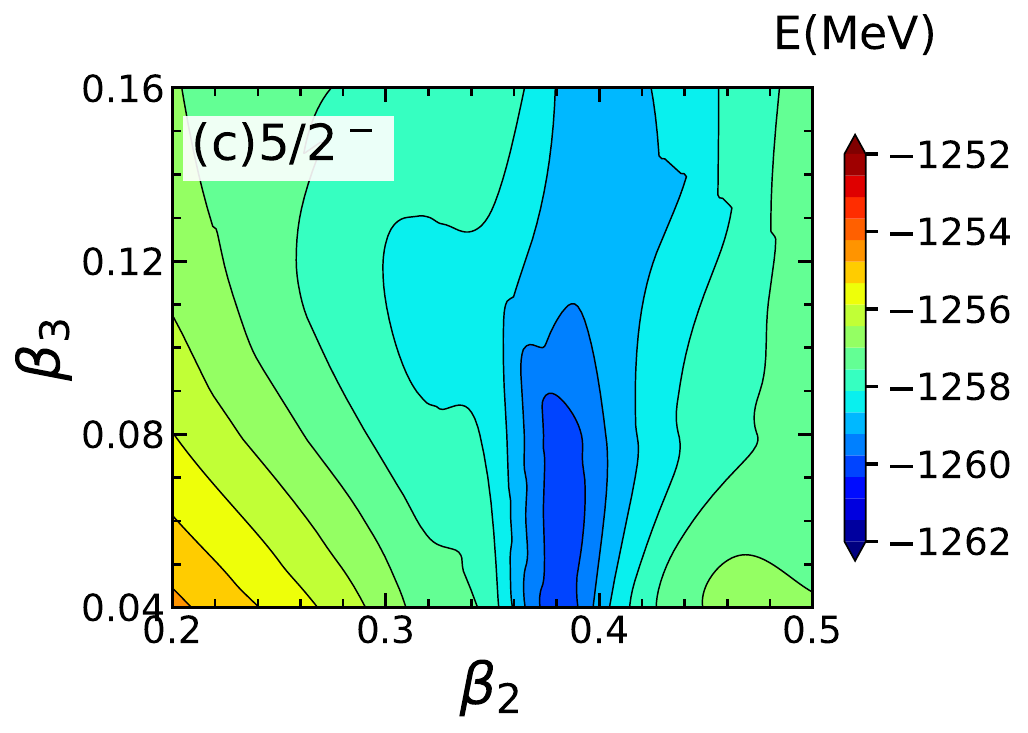}
\caption{The energy surfaces of $^{153}$Eu on the quadrupole--octupole deformation plane $(\beta_2,\beta_3)$ for configurations obtained at different levels of calculation with the PC-PK1 energy density functional. Panel (a) shows the mean-field results, while panels (b) and (c) display the energy surfaces after projection onto the correct nucleon numbers and spin-parity quantum numbers $J^{\pi}=5/2^{+}$ and $5/2^{-}$, respectively.
}
 \label{fig:PES}
 \end{figure}

Figure~\ref{fig:PES} displays energy surfaces of $^{153}$Eu on the quadrupole--octupole deformation plane $(\beta_2,\beta_3)$ obtained using different levels of MR-CDFT calculations. Fig.~\ref{fig:PES}(a) shows that the energy minimum of mean-field configurations is located around $(\beta_2=0.32, \beta_3=0.0)$, with softness towards the $\beta_3$ direction. After projection onto the correct nucleon numbers and spin-parity
$J^\pi=5/2^+$, the energy surface is rather
soft along both the $\beta_2$ and $\beta_3$ directions around the energy minimum. Two competing
low-energy regions can be identified: one around
$(\beta_2,\beta_3)\approx(0.30,0.05)$ and the other around
$(\beta_2,\beta_3)\approx(0.40,0.07$--$0.09)$. For both cases, the $\beta_2$ values are larger than the $\beta_2=0.249$ derived from the microscopic-macroscopic model \cite{Moller:2016}. In contrast, the $\beta_3=0.095$~\cite{Flambaum2025} extracted indirectly from data is more or less consistent with our prediction. The latter appears slightly deeper and more extended, indicating that the $5/2^+$ state favors a prolate quadrupole deformation with appreciable octupole deformation. The
broad low-energy valley connecting these two regions suggests significant quadrupole--octupole shape fluctuations and potential configuration mixing between different intrinsic shapes, as shown in Fig.~\ref{fig:PES}(b). For the $J^\pi=5/2^-$ state, the projected energy surface exhibits a more localized minimum around
$(\beta_2,\beta_3)\approx(0.39,0.07$--$0.08)$. Compared with the $5/2^+$ state, the low-energy valley is narrower in the $\beta_2$ direction and is more clearly centered at a larger prolate deformation, as shown in Fig.~\ref{fig:PES}(c). The surface remains moderately soft along the $\beta_3$ direction, but the minimum is more concentrated, suggesting a better-defined prolate--octupole intrinsic configuration for the negative-parity state. Overall, both states favor octupole-deformed shapes, while the $5/2^+$ state shows stronger shape coexistence or shape mixing and the $5/2^-$ state is dominated by a more localized prolate--octupole configuration.

 \begin{figure}[h]
 \centering
\includegraphics[width=\columnwidth]{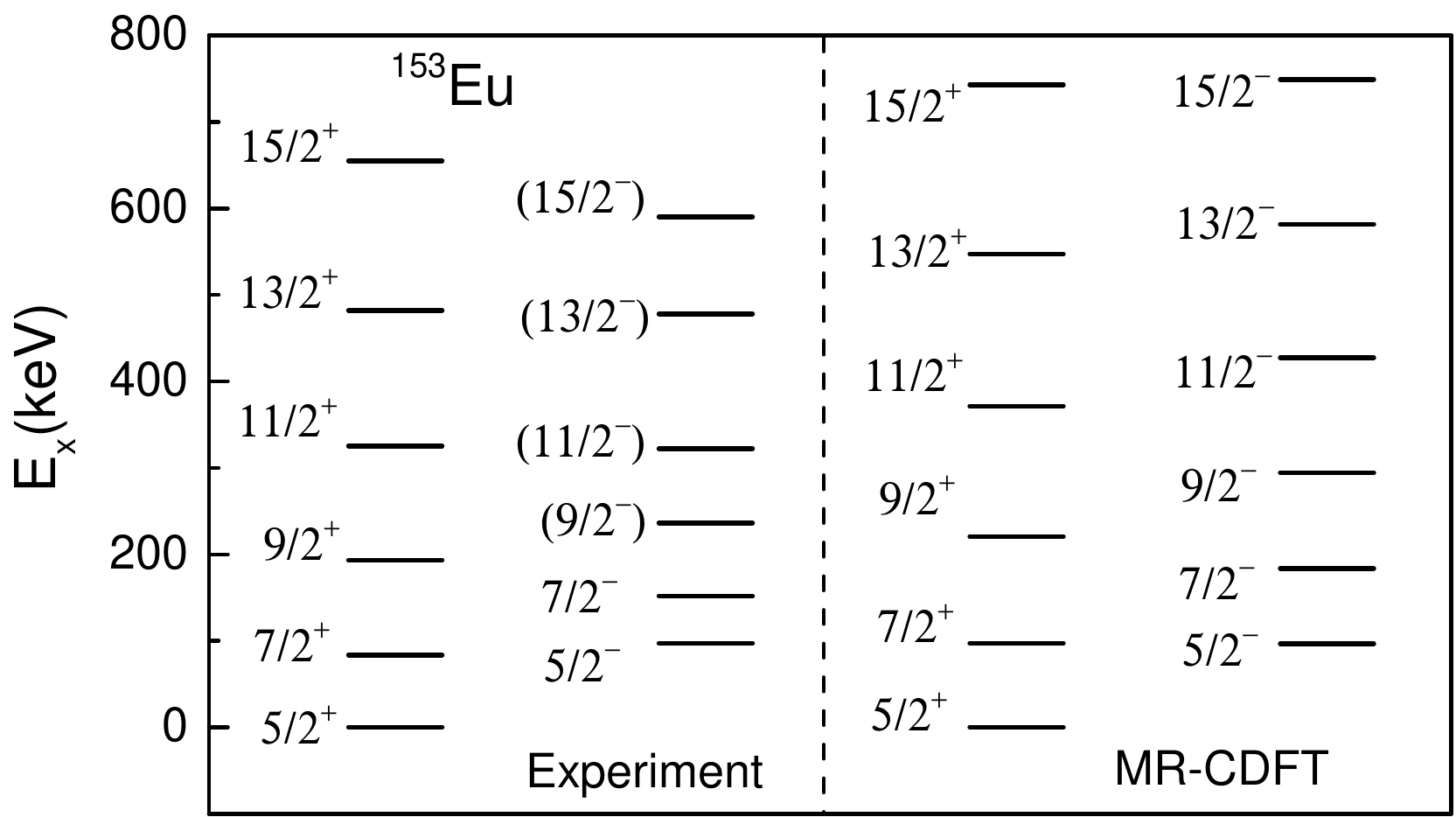}
\caption{The energy spectrum of low-lying states in $^{153}$Eu from the MR-CDFT calculations, in comparison with available experimental data \cite{NNDC}.}
 \label{fig:Eu153_level}
 \end{figure}

 \begin{figure}
 \centering
\includegraphics[width=0.8\columnwidth]{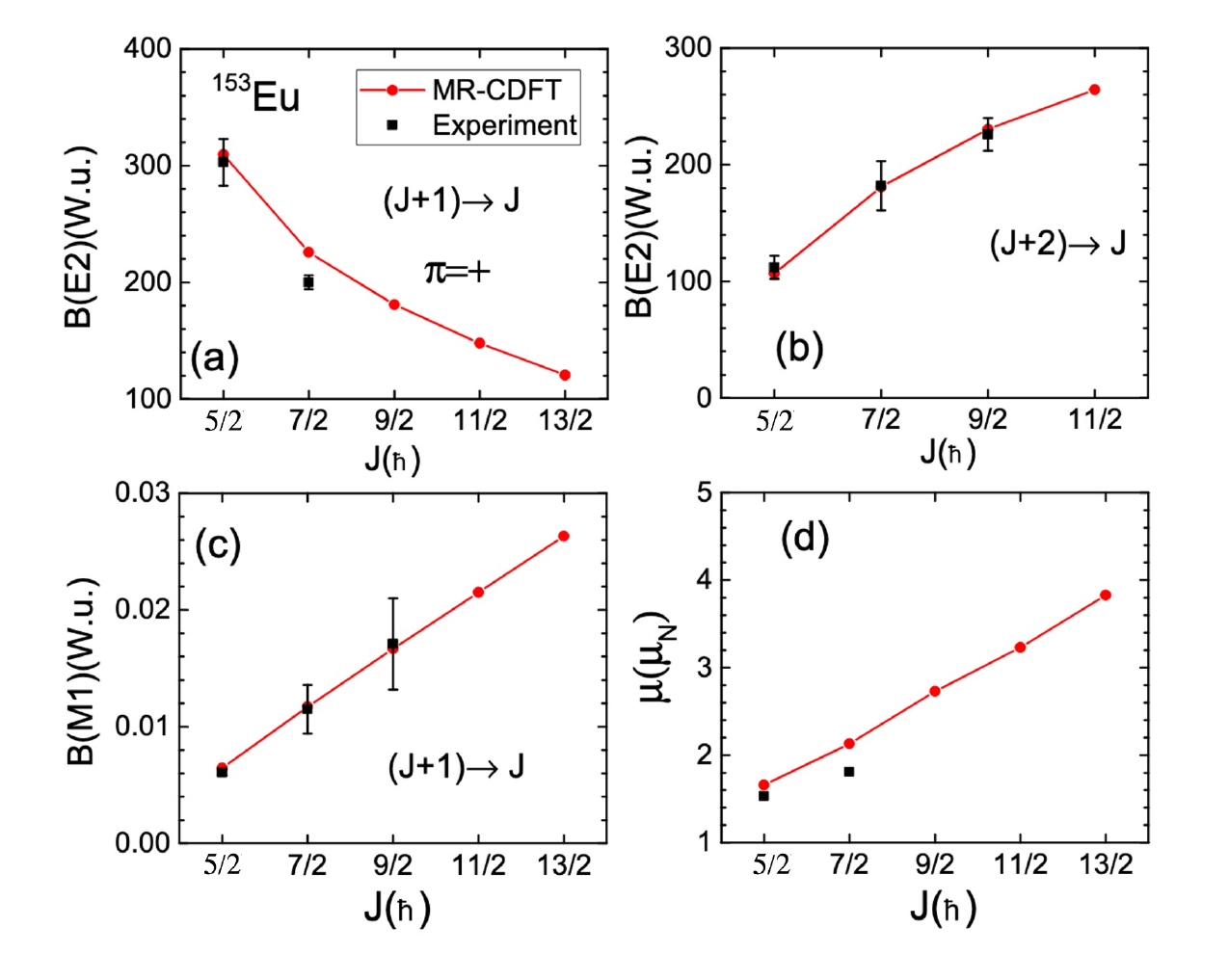}
\caption{Comparison between the calculated and measured electromagnetic observables of $^{153}$Eu. 
Panels (a, b) show the electric quadrupole transition strengths $B(E2)$ for both 
$(J+1)^+ \rightarrow J^+$ and $(J+2)^+ \rightarrow J^+$ transitions, respectively. 
Panel~(c) displays the magnetic dipole transition strengths $B(M1)$ for 
$(J+1)^+\rightarrow J^+$, and panel~(d) shows the magnetic dipole moments  $\mu(J^+)$ of the positive-parity states.  
}
 \label{fig:Eu153_E2_M1}
 \end{figure}
 
  \begin{figure}
 \centering
\includegraphics[width=\columnwidth]{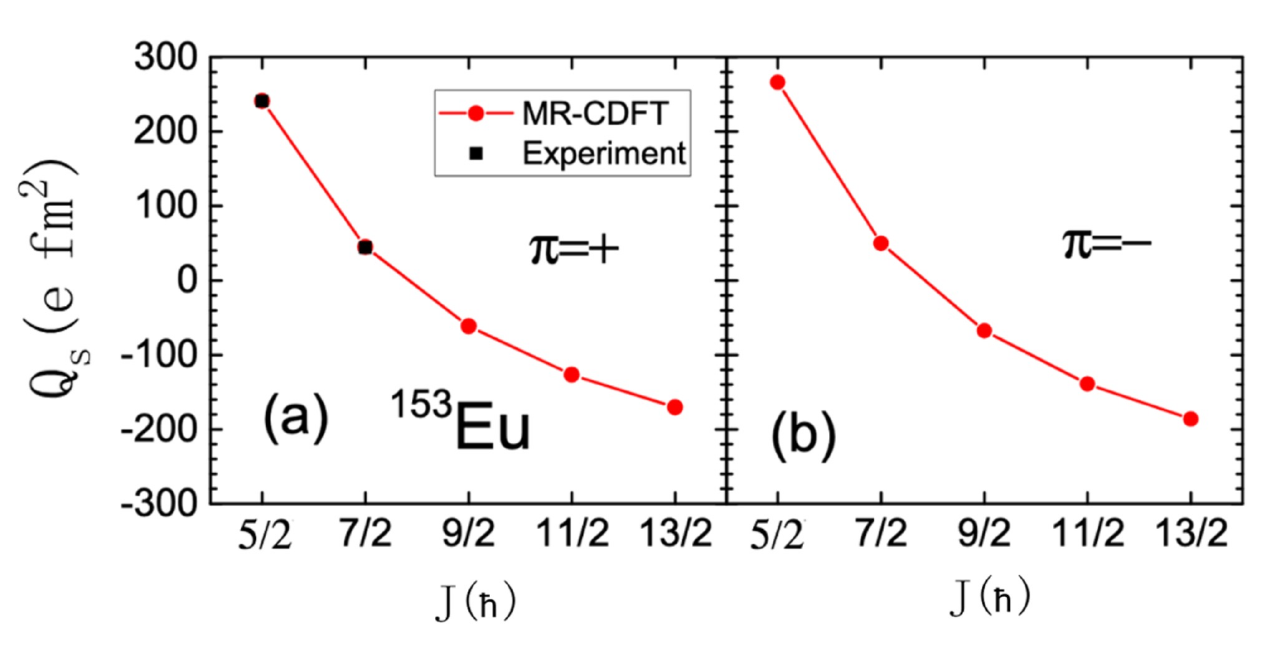}
\caption{Spectroscopic quadrupole moments $Q_s$ of the low-lying states of $^{153}\mathrm{Eu}$ from the MR-CDFT calculation, in comparison with available experimental data~\cite{NNDC}.
}
 \label{fig:Eu153_Qs}
 \end{figure}

The wavefunctions of nuclear states, $\ket{\Psi_k}$, were constructed by mixing the wave functions of mean-field states on the $(\beta_2,\beta_3)$ deformation plane, with the mixing weights determined variationally. This procedure leads to the Hill--Wheeler--Griffin equation~\cite{Ring:1980}, whose solutions provide the energies and mixing weights of the nuclear states $\ket{\Psi_k}$. These wavefunctions incorporate beyond-mean-field correlations associated with symmetry restoration, as well as both static and dynamical quadrupole--octupole deformations. Therefore we note that our calculation of the Schiff moment includes beyond-mean-field effects, which have been found to make a significant contribution to the Schiff moments of octupole-deformed nuclei~\cite{Zhou:2025_short}.

With these wave functions, one can straightforwardly evaluate observables of interest, including electric multipole transition strengths, magnetic dipole transition strengths, and magnetic dipole moments, which provide important benchmarks for validating the MR-CDFT calculations of Schiff moments \cite{Zhou:2025_Long}. Since the full single-particle basis is employed, there is no need to introduce effective charges or effective $g$ factors for neutrons and protons. This is an important feature that distinguishes our beyond-mean-field method from valence-space shell-model approaches.

Figure~\ref{fig:Eu153_level} displays the predicted energy spectrum of $^{153}$Eu obtained from the MR-CDFT calculation, in comparison with the corresponding experimental data. In addition, the predicted electric quadrupole and magnetic dipole transition strengths are shown in Fig.~\ref{fig:Eu153_E2_M1}, while the electric dipole (E1) and octupole (E3) transition strengths are compared with available data in Tables~\ref{tab:BE1} and \ref{tab:BE3}, respectively. The magnetic dipole moment of the first excited state $5/2^-$ is predicted to be 3.88 $\mu_N$, slightly larger than the experimental value 3.22(23) $\mu_N$~\cite{NNDC}.  The spectroscopic quadrupole moments are presented in Fig.~\ref{fig:Eu153_Qs}. Both the energy spectrum and the electromagnetic observables of $^{153}$Eu are in reasonable agreement with the available data.

Finally, the Schiff moment of $^{153}$Eu was calculated with $\eMax=12$ and the PC-PK1 parameter set, leading to Equation \ref{eq:schiff_calc} in the main text.

\section*{A2. Calculation of $\vec{\mathcal{W}}_\mrm{NSM}$}

$\vec{\mathcal{W}}_\mrm{NSM}$ is proportional to the derivative of the electron density at the position of the nucleus \cite{Dzuba02,Skripnikov20,Abe20,Hubert22,Chen24}. Therefore, it is necessary to employ all-electron relativistic electronic-structure methods for calculations of $\mathcal{W}_\mrm{NSM}$. We performed relativistic CASSCF calculations using our recent implementation of the two-component CASSCF method \cite{Wang2026} in the PySCF program package \cite{PySCF}. We used the exact two-component (X2C) Hamiltonian \cite{Dyall97,Kutzelnigg05,Ilias07,Liu09} with atomic mean-field spin-orbit integrals based on the Dirac-Coulomb-Breit Hamiltonian (the X2CAMF scheme) \cite{Liu18,Zhang22} in the present calculations.
The X2CAMF calculations include spin-orbit coupling variationally in the molecular spinors. We employed the X2CAMF program\cite{Zhang22,x2camf_git} to generate the relativistic effective one-electron integrals in the X2CAMF scheme and used the \texttt{socutils} module\cite{socutils_git} to interface the integrals into the PySCF program package. We carried out the subsequent RAS-CI calculations using the Dice program\cite{Sharma17_SHCI,Wang23_relSHCI}.
The X2C property integrals for $\vec{\mathcal{W}}_\mrm{NSM}$ and electric dipole moments were evaluated using the CFOUR program package \cite{CFOURfull,Matthews2020a}. We combined them with the CASSCF and CAS-CI density matrices and transition density matrices to obtain the $\vec{\mathcal{W}}_\mrm{NSM}$ and $\vec{d}_{00}$ values. 

In the CASSCF calculations, we distributed six Eu $4f$ electrons in the fourteen $4f$ spinors and performed state averaging over all states in the $4f^6$ manifold.  The subsequent RAS-CI calculations further included the O $2p$ spinors into the active space and allowed single excitations from the O $2p$ spinors into the Eu $4f$ spinors to account for the contributions to the $^7F_0\!\rightarrow{}^5\!D_0$ transition dipole moments from the ligand-to-metal charge transfer. The resulting RAS-CI active spaces comprise twelve electrons in twenty spinors for EuO$^+$ and Eu(OH)$^{2+}$, eighteen electrons in twenty-six spinors for Eu(OH)$_2^+$, and twenty-four electrons in thirty-two spinors for Eu(OH)$_3$. The calculations of Eu$^{3+}$ ions in uniform external electric fields
used CASCI calculations within only the CASSCF active space comprising six $4f$ electrons in fourteen $4f$ spinors.  

An even-tempered ``ETB0'' basis set for Eu, constructed and reported in Ref. \cite{Chen24} with enhanced flexibility in the core region to enable accurate calculations of the $^{151}$Eu nuclear Schiff moment sensitivity factors, was combined with the uncontracted cc-pVTZ basis sets for O and H in all the calculations presented here.

Two families of model systems were studied: (i) bare Eu$^{3+}$ ion in a uniform external electric field, in which the electric fields induce nonzero $\vec{d}_{00}$ and $\vec{\mathcal{W}}_{\mathrm{NSM}}$ in a way similar to the ligands in the model molecular systems; and (ii) a series of Eu--oxide (\ce{EuO+}) and Eu--hydroxide [\ce{Eu(OH)_x^{3-x}}] complexes that progressively approximate the crystal environment of an Eu$^{3+}$ ion doped into the Y$_2$SiO$_5$ crystal at the Y1 site with coordination number 7 (henceforth referred to as ``Eu:YSO''). We used the YSO crystal structure from the Materials Project \cite{Jain13_MP,MP_mp3520}.  The coordinates for the seven oxygen atoms on this Y1 site are given in Table~\ref{tab:Opos}. The model Eu-hydroxide complexes are
built by placing the oxygen atoms in subsets of these positions, 
keeping Eu-O-H linear and the O--H distance fixed at $0.974$~\AA. The specifics of calculations for the model systems are as follows.

\textit{Eu$^{3+}$ in uniform electric fields.}
The atomic ion is placed in a static uniform electric field. 
We calculated $\mathcal{W}_\mrm{NSM}$ and $d_{00}$ for  Eu$^{3+}$ ions at 10 field strengths $\Esca=0.01 n$~a.u. with $n=1, 2, 3, \cdots 10$. 

\textit{EuO$^+$.}
We computed the diatomic EuO$^+$ molecular ions with five Eu--O bond lengths
near its equilibrium bond length $r_e \approx 1.738$~\AA{} (3.284 a.u.), including $r=1.693$, $1.746$, $1.799$, $1.852$, and $1.958$~\AA.

\textit{Eu(OH)$^{2+}$.}
We computed Eu(OH)$^{2+}$ molecular ions in the linear Eu--O--H arrangement with
Eu--O distances of $2.207$, $2.302$, $2.378$, and $2.620$ {\AA} from Table~\ref{tab:Opos}.

\textit{Eu(OH)$_2^+$ and Eu(OH)$_3$.}
We construct two- and three-ligand complexes by placing OH$^-$ groups at the
subsets of the O positions in Table~\ref{tab:Opos}. 
We report the computational results for twenty distinct two-ligand combinations of Eu(OH)$_2^+$ and
twenty-one distinct three-ligand combinations of Eu(OH)$_3$, for which the CASSCF calculations successfully converged.

\textit{Eu(OH)$_7^{4-}$.}
We obtain the $\vec{d}_{00}$ and $\vec{\mathcal{W}}_{\mathrm{NSM}}$ parameters
for the complex with the coordination of seven OH ligands by means of an additivity scheme,
combining the $\vec{d}_{00}$ and $\vec{\mathcal{W}}_{\mathrm{NSM}}$ vectors of 
the seven corresponding individual Eu(OH)$^{2+}$ systems along the corresponding Eu--O directions at the Eu--O distances found in Eu:YSO (Table~\ref{tab:Opos}). This yields
$d_{00}=7.0\times10^{-4}$~a.u.\ and
$\mathcal{W}_\mrm{NSM}=3.8\times10^{3}$~a.u., with $\vec{\mathcal{W}}_{\mathrm{NSM}}$
projecting $70\%$ of its norm onto $\vec{d}_{00}$.  This is the
result for Eu(OH)$_7^{4-}$ used in the main text.

We validated the accuracy of the additivity scheme by comparing the results for the Eu(OH)$_2^+$ ions and
the Eu(OH)$_3$ molecules obtained from
the additivity scheme with those obtained from the calculations.  
For each of the twenty Eu(OH)$_2^+$ ions
and twenty-one Eu(OH)$_3$ molecules,
the results using the additivity scheme
--- obtained by summing the corresponding Eu(OH)$^{2+}$ results
--- are compared to the results obtained from calculations of these molecular systems  in 
Tables~\ref{tab:euoh2_add} and~\ref{tab:euoh3_add}. 
The results obtained using the additivity scheme are in very good agreement 
with those obtained from calculations.
The median and maximum 
deviations amount to $4\%$ and $12\%$ for the Eu(OH)$_2^+$ ions,
and to $11\%$ and $16\%$ for the Eu(OH)$_3$ molecules.

Tables~\ref{tab:eu_field_mag}--\ref{tab:euoh3_add} summarize
the $d_{00}$ and $\mathcal{W}_\mrm{NSM}$ values together with their ratios for
all model systems studied here. For the low-symmetry complexes,  $\cos\theta\equiv\vec{d}_{00}\!\cdot\vec{\mathcal{W}}_{\mathrm{NSM}}/(d_{00}\mathcal{W}_\mrm{NSM})$
and the Cartesian components of $\vec{d}_{00}$ and
$\vec{\mathcal{W}}_{\mathrm{NSM}}$ in the molecular frame are also reported.  
The global phase of $\vec{d}_{00}$ is fixed by the convention
$\vec{d}_{00}\!\cdot\vec{\mathcal{W}}_{\mathrm{NSM}}>0$.  
The mean value of the $\mathcal{W}_\mrm{NSM}/d_{00}$ ratios across the sixty-one model systems is $6.4\times10^{6}$.

The calculations also confirm the alignment between the $\vec{\mathcal{W}}_\mrm{NSM}$ and $\vec{d}_{00}$ vectors. For Eu:YSO, we adopt the results for the most realistic model system, the Eu(OH)$_7^{4-}$ ion, which has $d_{00}$ equal to $7\times 10^{-4}$, $\mathcal{W}_\mrm{NSM}$ equal to 3842 a.u. and $\cos \theta = 0.70$. Scaling the calculated $d_{00}$ by the experimental value of $8\times 10^{-4}$ a.u. \cite{Chen2025}, and accounting for the projection of $\mathcal{W}_\mrm{NSM}$ along $\vec{d}_{00}$, we estimate $\mathcal{W}_\mrm{NSM} \cdot \hat{x} = 3027$ a.u. 
            
The Eu(OH)$_7^{4-}$ model system has a $\mathcal{W}_\mrm{NSM}/d_{00}$ ratio of $\eta = 5.5\times10^{6}$, which is 14\% smaller than the value of $\eta = 6.4\times10^{6}$ obtained from the full set of 61 model systems. The $\cos\theta$ value of 0.7 in Eu(OH)$_7^{4-}$ is also smaller than those found in the other model systems, where $\cos\theta$ is close to unity. We use the deviation of the values for Eu(OH)$_7^{4-}$ from the average of all the model systems, $(\delta \mathcal{W}/\mathcal{W})_\mrm{dev.} \approx 14\%$, as a measure of the model error. For our conservative estimate of $\vec{\mathcal{W}}_\mrm{NSM} \cdot \hat{x}$ in Eu:YSO, we assume $\cos\theta = 0.7$ as calculated for Eu(OH)$_7^{4-}$, even though this value is atypical among the model systems and arises due to an accidental cancellation between the contributions from the various Eu(OH)$^{2+}$ groups in Eu(OH)$_7^{4-}$.

The $4f$ electrons in Eu make only small contributions to $\mathcal{W}_\mrm{NSM}$. This is reflected in the insensitivity of the computed $\mathcal{W}_\mrm{NSM}$ values to the occupation of the active orbitals. For example, the closed-shell HF $\mathcal{W}_\mrm{NSM}$ values deviate from the CASSCF results by less than 1\% of the total value, even though the closed-shell determinant contributes only about 1\% to the CASSCF wave function. It has been reported previously (15), and is also observed in the present study, that dynamic correlation on top of the closed-shell HF wave function reduces the HF $\mathcal{W}_\mrm{NSM}$ value in EuO$^+$ ions by about 15\%. In Eu$^{3+}$ subject to uniform external electric fields, this reduction is smaller, amounting to approximately 5\%. Therefore, the use of CASSCF results likely overestimates $\vec{\mathcal{W}}_\mrm{NSM}\cdot\hat{x}$ by $(\delta \mathcal{W}/\mathcal{W})_\mrm{dyn. \ corr.} \lesssim 15\%$ because of the neglect of dynamic correlation.

The error due to the model and the error due to neglect of dynamic correlation have opposite signs and tend to cancel. Nevertheless, the quadrature sum of the fractional errors due to these two error sources, $\delta \mathcal{W}/\mathcal{W} \approx 20\%$, is used as a conservative estimate of the error in $\vec{\mathcal{W}}_\mrm{NSM} \cdot \hat{x}$.  

\begin{figure}
    \centering
    \includegraphics[width=\linewidth]{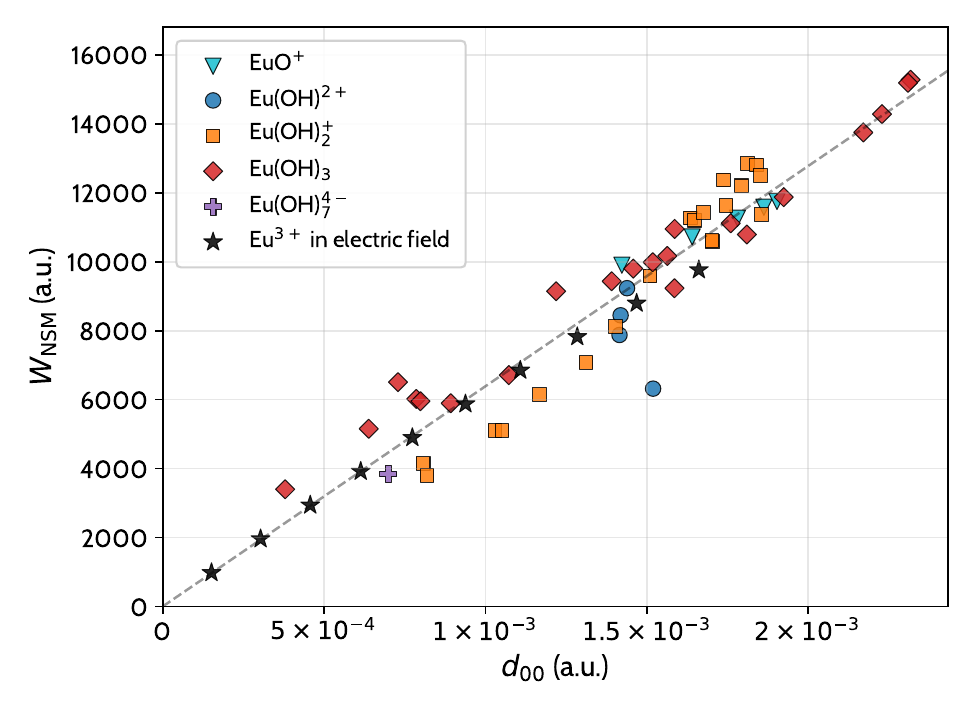}
    \caption{Correlation between calculated values of the Schiff moment interaction parameter, $\mathcal{W}_\mrm{NSM}$, and the transition dipole moment for the ${}^7F_0 - {}^5D_0$ optical transition, $d_{00}$, in a number of Eu-containing species. The dashed straight line has a slope of $6.4 \times 10^6$ a.u.}
    \label{fig:W_NSM_correlations}
\end{figure}

\section*{A3. Experiment}

\subsection*{Measurement method}

\begin{figure}[h!]
    \centering
    \includegraphics[width=0.5\linewidth]{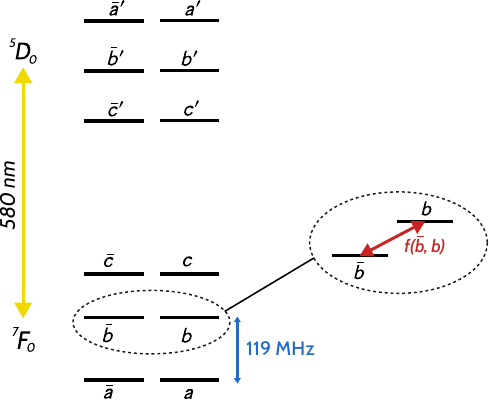}
    \caption{Energy levels in ${}^{153}$Eu:YSO. The electronic ground ${}^7F_0$ and excited ${}^5D_0$ states are connected by a 580.04 nm optical transition. Kramers-conjugate states, related by T-symmetry, are denoted as, e.g., $a, \bar{a}$. The $b-\bar{b}$ rf transition in the electronic ground state, in an applied magnetic field, is used for the measurement of the nuclear Schiff moment of $^{153}$Eu.}
    \label{fig:energy_levels}
\end{figure}

We used the apparatus and methods described in \cite{Nima2025} and \cite{Fan2026}. The measurements were made on a 0.01\% Eu-doped YSO crystal. The resonance frequency of the $b - \bar{b}$ transition, shown in Fig. \ref{fig:energy_levels}, was measured in the $\rho=\pm1$ sub-ensembles. No active or passive magnetic shielding was used in this measurement.

\begin{figure}[h!]
    \centering
    \includegraphics[width=1\linewidth]{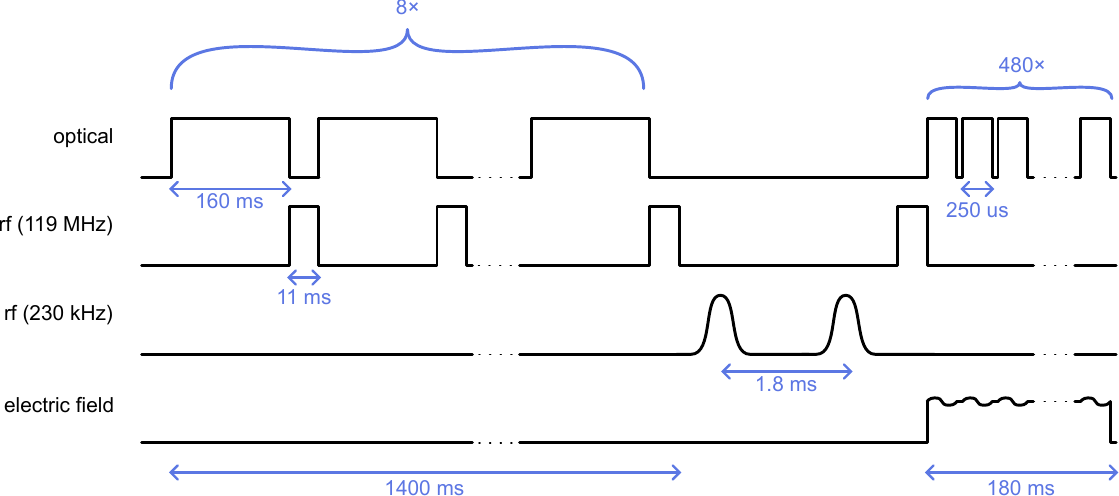}
    \caption{Timing diagram showing the sequence of laser, rf and electric field pulses used in each cycle of the experiment. Nominal parameters are shown for the sake of illustration: the pulse and timing parameters were varied outside these values during systematic error tests. The pulses shown in the ``rf (230 kHz)'' channel implement phasor Ramsey spectroscopy.}
    \label{fig:timing_diagram}
\end{figure}

Each measurement cycle involved the following steps, depicted in Fig.\ \ref{fig:timing_diagram}. First, the nuclear spin state was prepared in the ${}^7$F${}_0 \; \bar{b}$ state through optical pulses driving ${}^7$F${}_0 \;c, \bar{c} \rightarrow {}^5$D${}_0 \;b', \bar{b}'$, followed by rf pulses driving ${}^7$F${}_0 \;a, \bar{a} \rightarrow {}^5$D${}_0 \;c', \bar{c}'$, and then an adiabatic ${}^7$F${}_0 \;b \rightarrow {}^7$F${}_0 \;a$ rf sweep centered around 119 MHz. After repeating this cycle $N_\mrm{prep} = 8$ times, the ions were pumped into the ${}^7$F${}_0 \;\bar{b}$ state. The frequency of the ${}^7$F${}_0 \;a, \bar{a} \rightarrow {}^5$D${}_0 \;c', \bar{c}'$ laser used during this state-preparation sequence is denoted as $\nu_0$.

\begin{figure}[h!]
    \centering
    \includegraphics[width=1\linewidth]{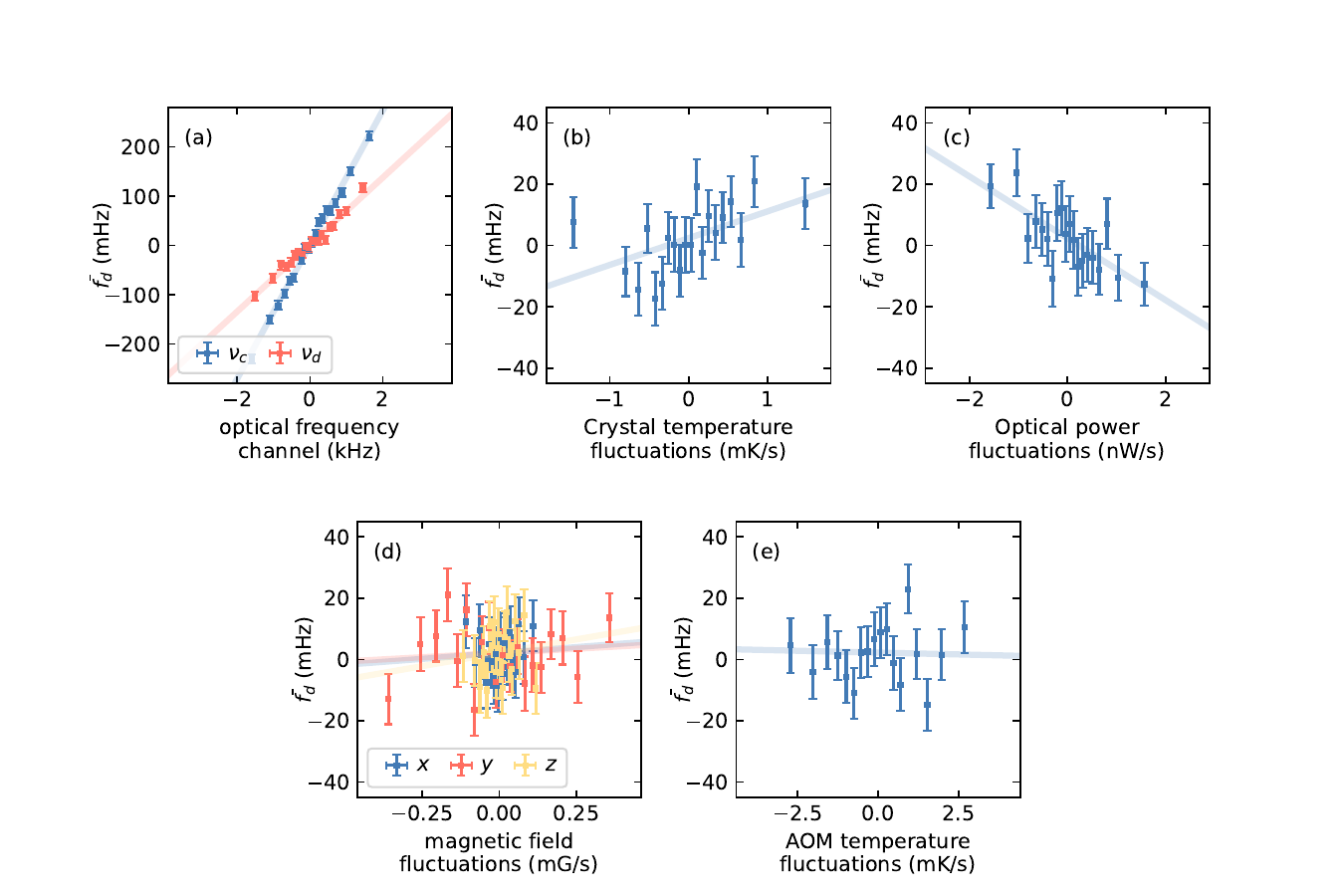}
    \caption{\textbf{Linear correlations.} Correlations between the measured $\bar{f}_d$ and a number of experiment parameters $x_i$ were analyzed. The most significant of these, which lead to the entries in Table \ref{tab:systematics}, are shown here.}
    \label{fig:corrections}
\end{figure}

Next the phasor Ramsey spectroscopy was performed. Two rf $\pi/2$ pulses with carrier frequency $f$, separated by time $\tau_R$, were used to drive the ${}^7$F${}_0 \;b - {}^7$F${}_0 \;\bar{b}$ transition. The primary dataset used $f$ = 230.47 kHz and $\tau_R \in \{1.8~\mrm{ms}, 2~\mrm{ms}\}$. Other values of $f$ and $\tau_R$ were used for consistency checks shown in Fig.\ \ref{fig:systematics}. The $\bar{b}-b$ population transfer due to the rf pulses was measured for various values of the relative phase, $\phi$, between the two pulses. The optical absorption signal -- proportional to the $b-\bar{b}$ population transfer -- was measured as a function of $\phi$ and fit to a sinusoid (see Fig.\ \ref{fig:ramsey}c). The phase offset $\phi_0$ obtained from this fit yields the resonance frequency $f_0$ through the relation \cite{Vutha2015}
\begin{equation}\label{eq:center_freq}
    f_0 = f -\frac{\phi_0}{2\pi \tau_R}.
\end{equation}

For state-detection, an adiabatic ${}^7$F${}_0 \;b \rightarrow {}^7$F${}_0 \;a$ rf sweep was used to transfer population to the ${}^7$F${}_0 \;a$ state. The population transferred was measured using frequency-modulated (FM) absorption on the ${}^7$F${}_0 \;a, \bar{a} \rightarrow {}^5$D${}_0 \;c, \bar{c}$ optical transition. A bias electric field $\Esca_\mrm{DC}$ (typically 33 V/cm) was applied along the $\hat{x}$ direction during the optical detection to distinguish the $\rho=\pm1$ populations. In this electric field, the $\rho = \pm1$ peaks appear at detunings $\nu - \nu_0 \approx \pm1$~MHz. The difference in the nuclear spin resonance frequencies measured for $\rho = \pm1$ yields $f_d$ in Equation (\ref{eq:fd}). 

The laser power transmitted through the crystal was measured on a photodiode. Another photodiode was used to monitor the laser power before the crystal. For the FM absorption measurement, the electric field was modulated with an amplitude $\mathcal{E}_\mathrm{AC}= 11$ V/cm at 4.0 kHz, and the demodulated absorption signal $a(\nu)$ was measured as a function of optical detuning $\nu$. Due to imperfections of the acousto-optic modulator (AOM) used to control the laser frequency and power, the optical power used for the FM measurements had a detuning-dependent background. To subtract this background, we reversed the sign of $\Esca_\mrm{DC}$ between each measurement, to exchange the $\rho= \pm1$ sub-ensembles, and measured $f_d(\pm \Esca_\mrm{DC})$. The average of $f_d(\pm \Esca_\mrm{DC})$, denoted as $\bar{f}_d$, has reduced dependence on the AOM background since the electric field reversal situates the $\rho = \pm1$ peaks at the same nominal detuning and background.

\subsection*{Systematic error analysis}
Systematic tests were conducted by measuring $\bar{f}_d$ at a variety of different values for experiment parameters. Experiment parameters were logged synchronously with the nuclear spin measurements. We studied the dependence of $\bar{f}_d$ on both the static values as well as the time derivatives of these parameters, since the nuclear spin resonance frequencies $f_0(\rho = \pm1)$ were measured non-simultaneously, in consecutive measurement cycles. 

For parameters where we observed correlations between the measured $\bar{f}_d$ and an experiment parameter $x_i$, corrections were applied as $\bar{f}_{d,\mrm{corrected}} = \bar{f}_{d,\mrm{measured}} - \frac{\partial \bar{f}_d}{\delta x_i} \avg{ x_i }$. Out of the large number of parameters studied during the tests, the following were significant.

\begin{enumerate}
    \item \textit{Optical spectrum correlations --} 
During the $\bar{f}_d$ measurement, absorption signals are measured at four detunings: $\nu(\rho=+1,\mathcal{E}_\mathrm{DC}>0)$, $\nu(\rho=-1,\mathcal{E}_\mathrm{DC}>0)$, $\nu(\rho=+1,\mathcal{E}_\mathrm{DC}<0)$, $\nu(\rho=-1,\mathcal{E}_\mathrm{DC}<0)$. 
There are two residual sources of imperfect subtraction of the AOM background: a) offset drift in the output of the high-voltage amplifier that applies $\Esca_\mrm{DC}$ during the time between the measurement of the $\pm \Esca_\mrm{DC}$ absorption features, which lead to the $\rho = \pm1$ lines appearing at slightly different detunings; b) laser frequency shifts during the time between $\pm \Esca_\mrm{DC}$ absorption measurements, that shift the $\rho = \pm1$ lines relative to the AOM background. Electric field offset drifts lead to a dependence of $\bar{f}_d$ on the optical detuning difference between the $\rho = \pm1$ absorption lines, defined as
\begin{equation}
\begin{split}
\nu_d  & = \frac{1}{2}\left\{ \nu(\rho=+1,\mathcal{E}_\mathrm{DC}>0)-\nu(\rho=-1,\mathcal{E}_\mathrm{DC}>0)\right\}\\
& + \frac{1}{2}\left\{\nu(\rho=+1,\mathcal{E}_\mathrm{DC}<0)-\nu(\rho=-1,\mathcal{E}_\mathrm{DC}<0)\right\}.
\end{split}
\end{equation}
Laser frequency drifts lead to a dependence on \begin{equation}
\begin{split}
\nu_c & 
= \frac{1}{2}\left\{\nu(\rho=+1,\mathcal{E}_\mathrm{DC}>0) 
+\nu(\rho=-1,\mathcal{E}_\mathrm{DC}>0)\right\}\\
& -\frac{1}{2}\left\{\nu(\rho=-1,\mathcal{E}_\mathrm{DC}<0)
+\nu(\rho=+1,\mathcal{E}_\mathrm{DC}<0)\right\}.
\end{split}
\end{equation}
If the AOM background were canceled perfectly, both $\nu_c$ and $\nu_d$ would be zero.We extracted $\nu_c$ and $\nu_d$ from the experimental dataset and studied their correlations with $\bar{f}_d$. We find $\avg{\nu_c} = 2.9 \pm 2.2$ Hz and $\avg{\nu_d} = 0.7 \pm 2$ Hz. The measured correlations with $\bar{f}_d$ are $\partial \bar{f}_d/\partial \nu_c = 136.2$ $\mu$Hz/Hz and $\partial \bar{f}_d/\partial \nu_d = 68$ $\mu$Hz/Hz (Fig.\ \ref{fig:corrections}a), leading to the corrections shown in Table \ref{tab:systematics}. To conservatively account for possible dependence between these two parameters, we summed their systematic uncertainties rather than combining them in quadrature.

Systematic errors due to this effect can be reduced in future measurements using a higher-bandwidth modulator, improvements to the laser stabilization method and using a shorter interval between $\pm \Esca_\mrm{DC}$ reversals.

\item \textit{Crystal temperature fluctuations --}
The temperature of the Eu:YSO crystal was monitored with a diode thermometer. The correlation between the time derivative of temperature fluctuations and $\bar{f}_d$ was $\partial \bar{f}_d / \partial (\partial_t T_\mathrm{xtl} ) = 8.7$ $\mu$Hz/($\mu$K/s), and the dataset had $\avg{\partial_t T_\mathrm{xtl}} = -1.1\pm7.3$ $\mu$K/s.

These temperature fluctuations were mainly produced by rf heating of metal components in the vicinity of the crystal. This source of error and uncertainty can be reduced in future measurements by rerouting the rf lines and improving the temperature stabilization of the crystal.

\item \textit{Laser power fluctuations --}
The laser power delivered to the crystal during the optical detection step was measured with a monitor photodiode and actively stabilized. However, residual laser power fluctuations between the $\rho = \pm1$ measurements can potentially lead to differential shifts between the measured spin resonance frequencies $f_0(\rho = \pm1)$. The dataset showed $\partial \bar{f}_d/\partial (\partial_tP_\mathrm{det}) = -10$ $\mu$Hz/(pW/s) and $\avg{\ \partial_tP_\mathrm{det} } = -1 \pm 2$ pW/s. 

Systematic effects due to laser power fluctuation can be suppressed in future measurements by improvements to the stabilization system and using a shorter interval between $\pm \Esca_\mrm{DC}$ reversals.

\item \textit{Magnetic field correlations --}
The magnetic field outside the experiment vacuum system was logged with a 3-axis fluxgate magnetometer. No correlations were observed with the magnetic field values, despite the absence of magnetic shielding, indicating that the comagnetometry method works accurately in practice. We conservatively included a systematic error due to correlations with the time derivative of the magnetic fields, which were measured to be $\partial \bar{f}_d/\partial(\partial_t\mathcal{B}_x) = 7.68$ nHz/(nG/s), $\partial \bar{f}_d/\partial(\partial_t\mathcal{B}_y) = 5.55$ nHz/(nG/s) and $\partial \bar{f}_d/\partial(\partial_t\mathcal{B}_z) = 17.48$ nHz/(nG/s). The derivatives of magnetic field fluctuations in the dataset were $\avg{\partial_t\mathcal{B}_x} = -30 \pm 150$ nG/s, $\avg{\partial_t\mathcal{B}_y} = -20 \pm 490$ nG/s, $\avg{\partial_t\mathcal{B}_z} = 10 \pm 160$ nG/s, consistent with zero. No correction was applied, but a systematic uncertainty was included in order to be conservative.

\item \textit{AOM temperature fluctuations --}
The AOM used to modulate the laser frequency and laser power during state-preparation and state-detection was temperature-controlled to suppress variations in its frequency response. The measured correlation of AOM temperature changes with $\bar{f}_d$ was $\partial \bar{f}_d/\partial(\partial_tT_\mathrm{AOM}) = -0.24$ $\mu$Hz/($\mu$K/s), and the dataset had $\avg{ \partial_tT_\mathrm{AOM} } = -11 \pm 6$ $\mu$K/s. This leads to the negligible correction and small uncertainty in Table \ref{fig:systematics}.
\end{enumerate}

All the systematic effects discussed here are technical in nature, suggesting that there is no fundamental barrier to a $\mu$Hz-level measurement of $\bar{f}_d$ in a crystal. 

\section*{A4. High-energy physics implications}
The isovector coupling parameter $\bar{g}_\pi^{(1)}$ is related to the energy scale of new physics that breaks T-symmetry, $M_{\cancel{T}}$, through the relation \cite{deVries2013}
\begin{equation}
\bar{g}_\pi^{(1)} \sim \frac{\bar{\delta}_3}{F_\pi} \frac{m_\pi^2 M_{QCD}}{M_{\cancel{T}}^2},
\end{equation}
where $m_\pi$ = 140 MeV is the pion mass, $F_\pi$ = 186 MeV is the pion decay constant, $M_{QCD} = 2 \pi F_\pi$ = 1.2 GeV, and $\bar{\delta}_3 \sim \frac{g_s}{4\pi} = 0.34$.
(Note that the quantity called $\bar{g}_1$ in Ref.\ \cite{deVries2013} is related to $\bar{g}_\pi^{(1)}$that appears in Equation \ref{eq:schiff_calc} as $\bar{g}_1 \approx F_\pi \bar{g}_\pi^{(1)}$.) Using this relation, the bound $|g_\pi^{(1)}| \leq 4. 6\times 10^{-10}$ leads to $M_{\cancel{T}} \gtrsim 9.7$ TeV.

\newpage
\subsection*{Appendix: Tables}

 \begin{table}[h!]
\centering
    \tabcolsep=6pt
    \caption{Electric dipole transition strengths $B(E1)$ (in Weisskopf units) of $^{153}\mathrm{Eu}$  calculated with the MR-CDFT approach, in comparison with available experimental data \cite{NNDC}.
}
    ~\\
    \begin{tabular}{llcc}
      \hline \hline 
 $J_f\rightarrow J_i$  &   &  MR-CDFT & Experiment  \\ 
   \hline
  $5/2^-\rightarrow 5/2^+$  &   & $4.2\times 10^{-4}$ & $9.8\times 10^{-4}(+9-7)$  \\  
$7/2^-\rightarrow 5/2^+$  &   & $1.1\times 10^{-4}$ & $3.0\times10^{-5} (+32-14)$  \\  
  $9/2^-\rightarrow 7/2^+$  &   & $1.9\times 10^{-4}$ & -  \\ 
  $11/2^-\rightarrow 9/2^+$  &   & $2.3\times 10^{-4}$ & -  \\ 
\hline\hline 
    \end{tabular}
    \label{tab:BE1}
\end{table} 

  \begin{table}[h!]
\centering
    \tabcolsep=6pt
    \caption{Electric octupole transition strengths $B(E3)$ (in Weisskopf units) of $^{153}$Eu calculated with the MR-CDFT approach.
}
    ~\\
    \begin{tabular}{llc}
      \hline \hline 
 $J_f\rightarrow J_i$  &   &  MR-CDFT   \\ 
   \hline
$7/2^-\rightarrow 5/2^+$  &   & $3.8$  \\  
  $9/2^-\rightarrow 7/2^+$  &   & 0.7  \\ 
  $11/2^-\rightarrow 9/2^+$  &   & 0.01 \\ 
  $13/2^-\rightarrow 11/2^+$  &   & 0.13 \\   
\hline
$9/2^-\rightarrow 5/2^+$  &   & $0.97$  \\  
  $11/2^-\rightarrow 7/2^+$  &   & 1.8  \\ 
  $13/2^-\rightarrow 9/2^+$  &   & 2.4 \\ 
  \hline
  \hline
    \end{tabular}
    \label{tab:BE3}
\end{table}

\begin{table}[h!]
\centering
\begin{tabular}{cccccc}
\hline
site & $x$ & $y$ & $z$ & $r$ (a.u.) & $r$ (\AA) \\
\hline
O1 & $+2.923$ & $+1.903$ & $-2.286$ & $4.170$ & $2.207$ \\
O2 & $-2.346$ & $+1.244$ & $-3.498$ & $4.392$ & $2.324$ \\
O3 & $+0.376$ & $-3.297$ & $-2.912$ & $4.415$ & $2.336$ \\
O4 & $+1.250$ & $+0.393$ & $+4.774$ & $4.951$ & $2.620$ \\
O5 & $-3.771$ & $+0.393$ & $+2.411$ & $4.493$ & $2.378$ \\
O6 & $-0.199$ & $+4.332$ & $+1.137$ & $4.483$ & $2.372$ \\
O7 & $+2.347$ & $-3.210$ & $+1.764$ & $4.350$ & $2.302$ \\
\hline
\end{tabular}
\caption{
The coordinates of the seven oxygen atoms at the Y1 site of the YSO crystal with
the metal placed at the origin together with the metal-oxygen bond distances.} 
\label{tab:Opos}
\end{table}

\begin{table}[h!]
\centering
\begin{tabular}{lccc}
\hline
$\Esca$ (a.u.) & $d_{00}$ & $\mathcal{W}_\mrm{NSM}$ & $\mathcal{W}_\mrm{NSM}/d_{00}$ \\
\hline
$0.01$ & $1.513\times 10^{-4}$ & $980$ & $6.481\times 10^{6}$ \\
$0.02$ & $3.034\times 10^{-4}$ & $1961$ & $6.463\times 10^{6}$ \\
$0.03$ & $4.572\times 10^{-4}$ & $2940$ & $6.431\times 10^{6}$ \\
$0.04$ & $6.137\times 10^{-4}$ & $3919$ & $6.387\times 10^{6}$ \\
$0.05$ & $7.737\times 10^{-4}$ & $4898$ & $6.330\times 10^{6}$ \\
$0.06$ & $9.382\times 10^{-4}$ & $5874$ & $6.261\times 10^{6}$ \\
$0.07$ & $1.108\times 10^{-3}$ & $6850$ & $6.181\times 10^{6}$ \\
$0.08$ & $1.285\times 10^{-3}$ & $7824$ & $6.090\times 10^{6}$ \\
$0.09$ & $1.469\times 10^{-3}$ & $8795$ & $5.988\times 10^{6}$ \\
$0.10$ & $1.662\times 10^{-3}$ & $9764$ & $5.877\times 10^{6}$ \\
\hline
\end{tabular}
\caption{Computed $d_{00}$ and $\mathcal{W}_\mrm{NSM}$ values (a.u.) and their ratios for the bare Eu$^{3+}$ ion in a uniform external electric field of strength $\Esca$.}
\label{tab:eu_field_mag}
\end{table}

\begin{table}[h!]
\centering
\begin{tabular}{lccc}
\hline
$r$ (\AA) & $d_{00}$ & $\mathcal{W}_\mrm{NSM}$ & $\mathcal{W}_\mrm{NSM}/d_{00}$ \\
\hline
$1.693$ & $1.423\times 10^{-3}$ & $9900$ & $6.956\times 10^{6}$ \\
$1.746$ & $1.641\times 10^{-3}$ & $10720$ & $6.533\times 10^{6}$ \\
$1.799$ & $1.781\times 10^{-3}$ & $11260$ & $6.322\times 10^{6}$ \\
$1.852$ & $1.863\times 10^{-3}$ & $11579$ & $6.215\times 10^{6}$ \\
$1.958$ & $1.904\times 10^{-3}$ & $11749$ & $6.172\times 10^{6}$ \\
\hline
\end{tabular}
\caption{Computed $d_{00}$ and $\mathcal{W}_\mrm{NSM}$ values (a.u.) and their ratios for the EuO$^+$ molecular ions at five bond lengths $r$.}
\label{tab:euo_mag}
\end{table}

\begin{table}[h!]
\centering
\begin{tabular}{llccc}
\hline
site & $r$ (\AA) & $d_{00}$ & $\mathcal{W}_\mrm{NSM}$ & $\mathcal{W}_\mrm{NSM}/d_{00}$ \\
\hline
O1 & 2.207 & $1.439 \times 10^{-3}$ & 9233 & $6.416 \times 10^6$ \\
O2 & 2.324 & $1.417 \times 10^{-3}$ & 8276 & $5.841 \times 10^6$ \\
O3 & 2.336 & $1.416 \times 10^{-3}$ & 8182 & $5.778 \times 10^6$ \\
O4 & 2.620 & $1.520 \times 10^{-3}$ & 6319 & $4.158 \times 10^6$ \\
O5 & 2.378 & $1.416 \times 10^{-3}$ & 7875 & $5.563 \times 10^6$ \\
O6 & 2.372 & $1.416 \times 10^{-3}$ & 7914 & $5.591 \times 10^6$ \\
O7 & 2.302 & $1.419 \times 10^{-3}$ & 8446 & $5.952 \times 10^6$ \\
\hline
\end{tabular}
\caption{Computed $d_{00}$ and $\mathcal{W}_\mrm{NSM}$ values (a.u.) and their ratios for the linear Eu(OH)$^{2+}$ molecular ion at seven Eu--O distances $r$ corresponding to the inter-atomic distances of the  Eu:YSO Y1 site. The O-H bond lengths are fixed to 0.974 {\AA}. The site labels  correspond to the O sites in Table \ref{tab:Opos}.}
\label{tab:euoh_mag}
\end{table}

\begin{table}[h!]
\centering
\begin{tabular}{lcccc}
\hline
sites & $d_{00}$ & $\mathcal{W}_\mrm{NSM}$ & $\mathcal{W}_\mrm{NSM}/d_{00}$ & $\cos\theta$ \\
\hline
O$_{1,2}$ & $1.813\times 10^{-3}$ & $12852$ & $7.090\times 10^{6}$ & 1.00 \\
O$_{1,3}$ & $1.738\times 10^{-3}$ & $12370$ & $7.116\times 10^{6}$ & 1.00 \\
O$_{1,5}$ & $1.029\times 10^{-3}$ & $5108$ & $4.963\times 10^{6}$ & 0.99 \\
O$_{1,6}$ & $1.840\times 10^{-3}$ & $12807$ & $6.962\times 10^{6}$ & 1.00 \\
O$_{1,7}$ & $1.636\times 10^{-3}$ & $11269$ & $6.890\times 10^{6}$ & 1.00 \\
O$_{2,3}$ & $1.793\times 10^{-3}$ & $12203$ & $6.806\times 10^{6}$ & 1.00 \\
O$_{2,4}$ & $8.074\times 10^{-4}$ & $4144$ & $5.132\times 10^{6}$ & 0.96 \\
O$_{2,5}$ & $1.648\times 10^{-3}$ & $11202$ & $6.796\times 10^{6}$ & 1.00 \\
O$_{2,6}$ & $1.676\times 10^{-3}$ & $11424$ & $6.818\times 10^{6}$ & 1.00 \\
O$_{2,7}$ & $1.052\times 10^{-3}$ & $5099$ & $4.846\times 10^{6}$ & 1.00 \\
O$_{3,4}$ & $1.168\times 10^{-3}$ & $6144$ & $5.260\times 10^{6}$ & 0.99 \\
O$_{3,5}$ & $1.404\times 10^{-3}$ & $8115$ & $5.781\times 10^{6}$ & 1.00 \\
O$_{3,6}$ & $8.186\times 10^{-4}$ & $3794$ & $4.634\times 10^{6}$ & 1.00 \\
O$_{3,7}$ & $1.852\times 10^{-3}$ & $12498$ & $6.748\times 10^{6}$ & 1.00 \\
O$_{4,5}$ & $1.703\times 10^{-3}$ & $10583$ & $6.213\times 10^{6}$ & 1.00 \\
O$_{4,6}$ & $1.703\times 10^{-3}$ & $10610$ & $6.230\times 10^{6}$ & 1.00 \\
O$_{4,7}$ & $1.856\times 10^{-3}$ & $11369$ & $6.126\times 10^{6}$ & 1.00 \\
O$_{5,6}$ & $1.746\times 10^{-3}$ & $11632$ & $6.662\times 10^{6}$ & 1.00 \\
O$_{5,7}$ & $1.510\times 10^{-3}$ & $9587$ & $6.348\times 10^{6}$ & 1.00 \\
O$_{6,7}$ & $1.312\times 10^{-3}$ & $7082$ & $5.398\times 10^{6}$ & 1.00 \\
\hline
\end{tabular}
\caption{Computed $d_{00}$ and $\mathcal{W}_\mrm{NSM}$ values (a.u.), their ratios, and the cosine of the angle $\theta$ between $\vec{d}_{00}$ and $\vec{\mathcal{W}}_{\mathrm{NSM}}$ for the Eu(OH)$_2^+$ complexes. The label O$_{i,j}$ identifies the pair of O sites in Table~\ref{tab:Opos} used to build the complex.}
\label{tab:euoh2_mag}
\end{table}

\begin{table}[h!]
\centering
\begin{tabular}{lcccc}
\hline
sites & $d_{00}$ & $\mathcal{W}_\mrm{NSM}$ & $\mathcal{W}_\mrm{NSM}/d_{00}$ & $\cos\theta$ \\
\hline
O$_{1,2,3}$ & $2.318\times 10^{-3}$ & $15280$ & $6.593\times 10^{6}$ & 1.00 \\
O$_{1,2,4}$ & $1.219\times 10^{-3}$ & $9144$ & $7.500\times 10^{6}$ & 0.98 \\
O$_{1,2,5}$ & $1.564\times 10^{-3}$ & $10166$ & $6.502\times 10^{6}$ & 0.99 \\
O$_{1,2,6}$ & $2.310\times 10^{-3}$ & $15191$ & $6.577\times 10^{6}$ & 1.00 \\
O$_{1,2,7}$ & $1.458\times 10^{-3}$ & $9800$ & $6.721\times 10^{6}$ & 0.98 \\
O$_{1,3,4}$ & $1.391\times 10^{-3}$ & $9434$ & $6.780\times 10^{6}$ & 0.97 \\
O$_{1,3,5}$ & $7.857\times 10^{-4}$ & $6020$ & $7.662\times 10^{6}$ & 0.96 \\
O$_{1,3,6}$ & $1.586\times 10^{-3}$ & $10952$ & $6.903\times 10^{6}$ & 1.00 \\
O$_{1,3,7}$ & $2.229\times 10^{-3}$ & $14282$ & $6.407\times 10^{6}$ & 1.00 \\
O$_{1,5,7}$ & $6.385\times 10^{-4}$ & $5155$ & $8.074\times 10^{6}$ & 0.96 \\
O$_{1,6,7}$ & $1.760\times 10^{-3}$ & $11115$ & $6.314\times 10^{6}$ & 0.99 \\
O$_{2,3,4}$ & $7.293\times 10^{-4}$ & $6506$ & $8.922\times 10^{6}$ & 0.99 \\
O$_{2,3,5}$ & $1.924\times 10^{-3}$ & $11873$ & $6.169\times 10^{6}$ & 1.00 \\
O$_{2,3,6}$ & $1.519\times 10^{-3}$ & $9986$ & $6.576\times 10^{6}$ & 1.00 \\
O$_{2,5,6}$ & $2.171\times 10^{-3}$ & $13751$ & $6.333\times 10^{6}$ & 1.00 \\
O$_{2,5,7}$ & $1.073\times 10^{-3}$ & $6711$ & $6.256\times 10^{6}$ & 0.98 \\
O$_{2,6,7}$ & $3.795\times 10^{-4}$ & $3399$ & $8.956\times 10^{6}$ & 0.96 \\
O$_{3,5,6}$ & $8.927\times 10^{-4}$ & $5896$ & $6.604\times 10^{6}$ & 0.98 \\
O$_{3,5,7}$ & $1.810\times 10^{-3}$ & $10790$ & $5.960\times 10^{6}$ & 0.99 \\
O$_{3,6,7}$ & $7.987\times 10^{-4}$ & $5955$ & $7.456\times 10^{6}$ & 0.97 \\
O$_{5,6,7}$ & $1.586\times 10^{-3}$ & $9231$ & $5.820\times 10^{6}$ & 0.99 \\
\hline
\end{tabular}
\caption{Computed $d_{00}$ and $\mathcal{W}_\mrm{NSM}$ (a.u.) values, their ratios, and the cosine of the angle $\theta$ between $\vec{d}_{00}$ and $\vec{\mathcal{W}}_{\mathrm{NSM}}$ for the Eu(OH)$_3$ complexes.
The label O$_{i,j,k}$ identifies the triple of O sites in Table~\ref{tab:Opos} used to build the complex.}
\label{tab:euoh3_mag}
\end{table}

\begin{table}[h!]
\centering
{\small
\begin{tabular}{lrrrrrr}
\hline
sites & $d_x$ & $d_y$ & $d_z$ & $\mathcal{W}_x$ & $\mathcal{W}_y$ & $\mathcal{W}_z$ \\
\hline
O$_{1,2}$ & $2.466\!\cdot\!10^{-4}$ & $8.751\!\cdot\!10^{-4}$ & $-1.568\!\cdot\!10^{-3}$ & $2020$ & $6238$ & $-11054$ \\
O$_{1,3}$ & $9.561\!\cdot\!10^{-4}$ & $-2.899\!\cdot\!10^{-4}$ & $-1.422\!\cdot\!10^{-3}$ & $6950$ & $-1759$ & $-10081$ \\
O$_{1,5}$ & $-1.465\!\cdot\!10^{-4}$ & $1.014\!\cdot\!10^{-3}$ & $-9.536\!\cdot\!10^{-5}$ & $-28$ & $5022$ & $-931$ \\
O$_{1,6}$ & $8.016\!\cdot\!10^{-4}$ & $1.610\!\cdot\!10^{-3}$ & $-3.875\!\cdot\!10^{-4}$ & $5789$ & $11043$ & $-2927$ \\
O$_{1,7}$ & $1.590\!\cdot\!10^{-3}$ & $-3.188\!\cdot\!10^{-4}$ & $-2.149\!\cdot\!10^{-4}$ & $10972$ & $-1961$ & $-1665$ \\
O$_{2,3}$ & $-5.083\!\cdot\!10^{-4}$ & $-5.169\!\cdot\!10^{-4}$ & $-1.640\!\cdot\!10^{-3}$ & $-3470$ & $-3500$ & $-11164$ \\
O$_{2,4}$ & $-5.317\!\cdot\!10^{-4}$ & $6.013\!\cdot\!10^{-4}$ & $8.797\!\cdot\!10^{-5}$ & $-2906$ & $2890$ & $-614$ \\
O$_{2,5}$ & $-1.560\!\cdot\!10^{-3}$ & $4.259\!\cdot\!10^{-4}$ & $-3.194\!\cdot\!10^{-4}$ & $-10570$ & $2912$ & $-2298$ \\
O$_{2,6}$ & $-6.628\!\cdot\!10^{-4}$ & $1.403\!\cdot\!10^{-3}$ & $-6.316\!\cdot\!10^{-4}$ & $-4559$ & $9510$ & $-4392$ \\
O$_{2,7}$ & $1.897\!\cdot\!10^{-5}$ & $-8.067\!\cdot\!10^{-4}$ & $-6.753\!\cdot\!10^{-4}$ & $147$ & $-3960$ & $-3209$ \\
O$_{3,4}$ & $4.676\!\cdot\!10^{-4}$ & $-1.032\!\cdot\!10^{-3}$ & $2.849\!\cdot\!10^{-4}$ & $2288$ & $-5670$ & $604$ \\
O$_{3,5}$ & $-1.035\!\cdot\!10^{-3}$ & $-9.293\!\cdot\!10^{-4}$ & $-1.893\!\cdot\!10^{-4}$ & $-5905$ & $-5437$ & $-1190$ \\
O$_{3,6}$ & $7.397\!\cdot\!10^{-5}$ & $3.631\!\cdot\!10^{-4}$ & $-7.300\!\cdot\!10^{-4}$ & $353$ & $1537$ & $-3451$ \\
O$_{3,7}$ & $7.159\!\cdot\!10^{-4}$ & $-1.686\!\cdot\!10^{-3}$ & $-2.739\!\cdot\!10^{-4}$ & $4852$ & $-11376$ & $-1798$ \\
O$_{4,5}$ & $-7.044\!\cdot\!10^{-4}$ & $1.759\!\cdot\!10^{-4}$ & $1.541\!\cdot\!10^{-3}$ & $-4696$ & $1093$ & $9421$ \\
O$_{4,6}$ & $1.939\!\cdot\!10^{-4}$ & $1.172\!\cdot\!10^{-3}$ & $1.221\!\cdot\!10^{-3}$ & $1113$ & $7556$ & $7365$ \\
O$_{4,7}$ & $8.913\!\cdot\!10^{-4}$ & $-8.134\!\cdot\!10^{-4}$ & $1.410\!\cdot\!10^{-3}$ & $5550$ & $-5270$ & $8407$ \\
O$_{5,6}$ & $-9.713\!\cdot\!10^{-4}$ & $1.162\!\cdot\!10^{-3}$ & $8.693\!\cdot\!10^{-4}$ & $-6465$ & $7746$ & $5789$ \\
O$_{5,7}$ & $-3.453\!\cdot\!10^{-4}$ & $-8.521\!\cdot\!10^{-4}$ & $1.198\!\cdot\!10^{-3}$ & $-2000$ & $-5517$ & $7581$ \\
O$_{6,7}$ & $7.746\!\cdot\!10^{-4}$ & $3.037\!\cdot\!10^{-4}$ & $1.014\!\cdot\!10^{-3}$ & $4258$ & $1385$ & $5486$ \\
\hline
\end{tabular}
}
\caption{Cartesian components (a.u.) of $\vec{d}_{00}$ and $\vec{\mathcal{W}}_{\mathrm{NSM}}$ in the molecular frame for the Eu(OH)$_2^+$ complexes.  The global phase of $\vec{d}_{00}$ has been fixed by the convention $\vec{d}_{00}\!\cdot\vec{\mathcal{W}}_{\mathrm{NSM}}>0$.}
\label{tab:euoh2_comp}
\end{table}

\begin{table}[h!]
\centering
{\small
\begin{tabular}{lrrrrrr}
\hline
sites & $d_x$ & $d_y$ & $d_z$ & $\mathcal{W}_x$ & $\mathcal{W}_y$ & $\mathcal{W}_z$ \\
\hline
O$_{1,2,3}$ & $3.838\!\cdot\!10^{-4}$ & $3.003\!\cdot\!10^{-5}$ & $-2.286\!\cdot\!10^{-3}$ & $2720$ & $539$ & $-15026$ \\
O$_{1,2,4}$ & $3.970\!\cdot\!10^{-4}$ & $1.034\!\cdot\!10^{-3}$ & $-5.103\!\cdot\!10^{-4}$ & $3415$ & $6588$ & $-5343$ \\
O$_{1,2,5}$ & $-6.251\!\cdot\!10^{-4}$ & $1.152\!\cdot\!10^{-3}$ & $-8.533\!\cdot\!10^{-4}$ & $-3777$ & $6691$ & $-6658$ \\
O$_{1,2,6}$ & $1.505\!\cdot\!10^{-4}$ & $1.918\!\cdot\!10^{-3}$ & $-1.279\!\cdot\!10^{-3}$ & $1573$ & $12406$ & $-8623$ \\
O$_{1,2,7}$ & $8.448\!\cdot\!10^{-4}$ & $-2.164\!\cdot\!10^{-4}$ & $-1.169\!\cdot\!10^{-3}$ & $6310$ & $97$ & $-7498$ \\
O$_{1,3,4}$ & $1.305\!\cdot\!10^{-3}$ & $-3.328\!\cdot\!10^{-4}$ & $-3.505\!\cdot\!10^{-4}$ & $8279$ & $-1250$ & $-4346$ \\
O$_{1,3,5}$ & $-1.192\!\cdot\!10^{-4}$ & $-8.095\!\cdot\!10^{-6}$ & $-7.766\!\cdot\!10^{-4}$ & $633$ & $-947$ & $-5911$ \\
O$_{1,3,6}$ & $7.725\!\cdot\!10^{-4}$ & $7.510\!\cdot\!10^{-4}$ & $-1.164\!\cdot\!10^{-3}$ & $6057$ & $4911$ & $-7690$ \\
O$_{1,3,7}$ & $1.712\!\cdot\!10^{-3}$ & $-1.076\!\cdot\!10^{-3}$ & $-9.380\!\cdot\!10^{-4}$ & $10815$ & $-6830$ & $-6353$ \\
O$_{1,5,7}$ & $5.022\!\cdot\!10^{-4}$ & $-1.999\!\cdot\!10^{-6}$ & $3.942\!\cdot\!10^{-4}$ & $4478$ & $-1144$ & $2282$ \\
O$_{1,6,7}$ & $1.569\!\cdot\!10^{-3}$ & $7.340\!\cdot\!10^{-4}$ & $3.150\!\cdot\!10^{-4}$ & $10031$ & $4769$ & $425$ \\
O$_{2,3,4}$ & $-2.767\!\cdot\!10^{-4}$ & $-3.978\!\cdot\!10^{-4}$ & $-5.450\!\cdot\!10^{-4}$ & $-1984$ & $-2969$ & $-5439$ \\
O$_{2,3,5}$ & $-1.554\!\cdot\!10^{-3}$ & $-5.959\!\cdot\!10^{-4}$ & $-9.659\!\cdot\!10^{-4}$ & $-9330$ & $-2876$ & $-6757$ \\
O$_{2,3,6}$ & $-4.266\!\cdot\!10^{-4}$ & $5.624\!\cdot\!10^{-4}$ & $-1.345\!\cdot\!10^{-3}$ & $-3457$ & $3468$ & $-8703$ \\
O$_{2,5,6}$ & $-1.597\!\cdot\!10^{-3}$ & $1.469\!\cdot\!10^{-3}$ & $-8.075\!\cdot\!10^{-5}$ & $-10062$ & $9358$ & $-523$ \\
O$_{2,5,7}$ & $-8.472\!\cdot\!10^{-4}$ & $-6.573\!\cdot\!10^{-4}$ & $3.164\!\cdot\!10^{-5}$ & $-5828$ & $-3186$ & $963$ \\
O$_{2,6,7}$ & $8.670\!\cdot\!10^{-5}$ & $3.400\!\cdot\!10^{-4}$ & $-1.447\!\cdot\!10^{-4}$ & $-54$ & $3244$ & $-1012$ \\
O$_{3,5,6}$ & $-8.438\!\cdot\!10^{-4}$ & $2.658\!\cdot\!10^{-4}$ & $-1.193\!\cdot\!10^{-4}$ & $-5622$ & $1720$ & $443$ \\
O$_{3,5,7}$ & $-4.619\!\cdot\!10^{-4}$ & $-1.706\!\cdot\!10^{-3}$ & $3.914\!\cdot\!10^{-4}$ & $-1539$ & $-10445$ & $2223$ \\
O$_{3,6,7}$ & $7.037\!\cdot\!10^{-4}$ & $-3.771\!\cdot\!10^{-4}$ & $-2.040\!\cdot\!10^{-5}$ & $4465$ & $-3938$ & $142$ \\
O$_{5,6,7}$ & $-1.449\!\cdot\!10^{-4}$ & $2.706\!\cdot\!10^{-4}$ & $1.556\!\cdot\!10^{-3}$ & $-1816$ & $1552$ & $8917$ \\
\hline
\end{tabular}
}
\caption{Cartesian components (a.u.) of $\vec{d}_{00}$ and $\vec{\mathcal{W}}_{\mathrm{NSM}}$ in the molecular frame for the Eu(OH)$_3$ complexes.  The global phase of $\vec{d}_{00}$ has been fixed by the convention $\vec{d}_{00}\!\cdot\vec{\mathcal{W}}_{\mathrm{NSM}}>0$.}
\label{tab:euoh3_comp}
\end{table}

\begin{table}[h!]
\centering
{\small
\begin{tabular}{lcccccc}
\hline
sites & $d_{00, \mathrm{dir}}$ & $d_{00, \mathrm{add}}$ & $\Delta d_{00}$ (\%) & $\mathcal{W}_{\mrm{NSM}, \mathrm{dir}}$ & $\mathcal{W}_{\mrm{NSM},\mathrm{add}}$ & $\Delta\mathcal{W}_\mrm{NSM}$ (\%) \\
\hline
O$_{1,2}$ & $1.813\times 10^{-3}$ & $2.205\times 10^{-3}$ & $+21.6$ & $12852$ & $13532$ & $+5.3$ \\
O$_{1,3}$ & $1.738\times 10^{-3}$ & $2.100\times 10^{-3}$ & $+20.8$ & $12370$ & $12824$ & $+3.7$ \\
O$_{1,5}$ & $1.029\times 10^{-3}$ & $8.016\times 10^{-4}$ & $-22.1$ & $5108$ & $4976$ & $-2.6$ \\
O$_{1,6}$ & $1.840\times 10^{-3}$ & $2.276\times 10^{-3}$ & $+23.7$ & $12807$ & $13691$ & $+6.9$ \\
O$_{1,7}$ & $1.636\times 10^{-3}$ & $1.829\times 10^{-3}$ & $+11.8$ & $11269$ & $11330$ & $+0.5$ \\
O$_{2,3}$ & $1.793\times 10^{-3}$ & $2.258\times 10^{-3}$ & $+25.9$ & $12203$ & $13115$ & $+7.5$ \\
O$_{2,4}$ & $8.074\times 10^{-4}$ & $7.248\times 10^{-4}$ & $-10.2$ & $4144$ & $4044$ & $-2.4$ \\
O$_{2,5}$ & $1.648\times 10^{-3}$ & $2.049\times 10^{-3}$ & $+24.3$ & $11202$ & $11685$ & $+4.3$ \\
O$_{2,6}$ & $1.676\times 10^{-3}$ & $2.097\times 10^{-3}$ & $+25.2$ & $11424$ & $11987$ & $+4.9$ \\
O$_{2,7}$ & $1.052\times 10^{-3}$ & $8.509\times 10^{-4}$ & $-19.1$ & $5099$ & $5019$ & $-1.6$ \\
O$_{3,4}$ & $1.168\times 10^{-3}$ & $1.190\times 10^{-3}$ & $+1.8$ & $6144$ & $6102$ & $-0.7$ \\
O$_{3,5}$ & $1.404\times 10^{-3}$ & $1.430\times 10^{-3}$ & $+1.9$ & $8115$ & $8109$ & $-0.1$ \\
O$_{3,6}$ & $8.186\times 10^{-4}$ & $6.563\times 10^{-4}$ & $-19.8$ & $3794$ & $3739$ & $-1.4$ \\
O$_{3,7}$ & $1.852\times 10^{-3}$ & $2.313\times 10^{-3}$ & $+24.9$ & $12498$ & $13563$ & $+8.5$ \\
O$_{4,5}$ & $1.703\times 10^{-3}$ & $2.379\times 10^{-3}$ & $+39.7$ & $10583$ & $11534$ & $+9.0$ \\
O$_{4,6}$ & $1.703\times 10^{-3}$ & $2.377\times 10^{-3}$ & $+39.6$ & $10610$ & $11558$ & $+8.9$ \\
O$_{4,7}$ & $1.856\times 10^{-3}$ & $2.519\times 10^{-3}$ & $+35.7$ & $11369$ & $12699$ & $+11.7$ \\
O$_{5,6}$ & $1.746\times 10^{-3}$ & $2.246\times 10^{-3}$ & $+28.6$ & $11632$ & $12522$ & $+7.6$ \\
O$_{5,7}$ & $1.510\times 10^{-3}$ & $1.677\times 10^{-3}$ & $+11.1$ & $9587$ & $9667$ & $+0.8$ \\
O$_{6,7}$ & $1.312\times 10^{-3}$ & $1.213\times 10^{-3}$ & $-7.6$ & $7082$ & $7014$ & $-1.0$ \\
\hline
\end{tabular}
}
\caption{Validation of the additivity scheme using the Eu(OH)$_2^+$ molecular ions.  For each of the twenty two-ligand combinations, the $d_{00}$ and $\mathcal{W}_\mrm{NSM}$ values (a.u.) obtained directly from the calculations (denoted with subscripts ``dir'') are compared with the values obtained using the additivity scheme (denoted with subscripts ``add'') that sums the parameters of the two corresponding Eu(OH)$^{2+}$ complexes listed in Table~\ref{tab:Opos}.  The fourth and the last columns show the percentage deviations $\Delta X=(X_{\mathrm{add}}/X_{\mathrm{dir}}-1)\times 100\%$.}
\label{tab:euoh2_add}
\end{table}

\begin{table}[h!]
\centering
{\small
\begin{tabular}{lcccccc}
\hline
sites & $d_{00,\mathrm{dir}}$ & $d_{00,\mathrm{add}}$ & $\Delta d_{00}$ (\%) & $\mathcal{W}_{\mrm{NSM}, \mathrm{dir}}$ & $\mathcal{W}_{\mrm{NSM},\mathrm{add}}$ & $\Delta\mathcal{W}_\mrm{NSM}$ (\%) \\
\hline
O$_{1,2,3}$ & $2.318\times 10^{-3}$ & $2.878\times 10^{-3}$ & $+24.2$ & $15280$ & $17283$ & $+13.1$ \\
O$_{1,2,4}$ & $1.219\times 10^{-3}$ & $1.413\times 10^{-3}$ & $+15.9$ & $9144$ & $9699$ & $+6.1$ \\
O$_{1,2,5}$ & $1.564\times 10^{-3}$ & $1.902\times 10^{-3}$ & $+21.7$ & $10166$ & $11340$ & $+11.5$ \\
O$_{1,2,6}$ & $2.310\times 10^{-3}$ & $2.890\times 10^{-3}$ & $+25.1$ & $15191$ & $17257$ & $+13.6$ \\
O$_{1,2,7}$ & $1.458\times 10^{-3}$ & $1.684\times 10^{-3}$ & $+15.5$ & $9800$ & $10560$ & $+7.8$ \\
O$_{1,3,4}$ & $1.391\times 10^{-3}$ & $1.560\times 10^{-3}$ & $+12.1$ & $9434$ & $9891$ & $+4.9$ \\
O$_{1,3,5}$ & $7.857\times 10^{-4}$ & $1.005\times 10^{-3}$ & $+27.9$ & $6020$ & $6378$ & $+5.9$ \\
O$_{1,3,6}$ & $1.586\times 10^{-3}$ & $1.983\times 10^{-3}$ & $+25.0$ & $10952$ & $12287$ & $+12.2$ \\
O$_{1,3,7}$ & $2.229\times 10^{-3}$ & $2.647\times 10^{-3}$ & $+18.8$ & $14282$ & $15911$ & $+11.4$ \\
O$_{1,5,7}$ & $6.385\times 10^{-4}$ & $8.441\times 10^{-4}$ & $+32.2$ & $5155$ & $5290$ & $+2.6$ \\
O$_{1,6,7}$ & $1.760\times 10^{-3}$ & $1.976\times 10^{-3}$ & $+12.3$ & $11115$ & $12074$ & $+8.6$ \\
O$_{2,3,4}$ & $7.293\times 10^{-4}$ & $8.429\times 10^{-4}$ & $+15.6$ & $6506$ & $7076$ & $+8.8$ \\
O$_{2,3,5}$ & $1.924\times 10^{-3}$ & $2.306\times 10^{-3}$ & $+19.8$ & $11873$ & $13293$ & $+12.0$ \\
O$_{2,3,6}$ & $1.519\times 10^{-3}$ & $1.976\times 10^{-3}$ & $+30.1$ & $9986$ & $11465$ & $+14.8$ \\
O$_{2,5,6}$ & $2.171\times 10^{-3}$ & $2.760\times 10^{-3}$ & $+27.1$ & $13751$ & $15615$ & $+13.6$ \\
O$_{2,5,7}$ & $1.073\times 10^{-3}$ & $1.306\times 10^{-3}$ & $+21.8$ & $6711$ & $7299$ & $+8.8$ \\
O$_{2,6,7}$ & $3.795\times 10^{-4}$ & $7.503\times 10^{-4}$ & $+97.7$ & $3399$ & $3942$ & $+16.0$ \\
O$_{3,5,6}$ & $8.927\times 10^{-4}$ & $1.224\times 10^{-3}$ & $+37.2$ & $5896$ & $6697$ & $+13.6$ \\
O$_{3,5,7}$ & $1.810\times 10^{-3}$ & $2.044\times 10^{-3}$ & $+12.9$ & $10790$ & $11949$ & $+10.7$ \\
O$_{3,6,7}$ & $7.987\times 10^{-4}$ & $1.106\times 10^{-3}$ & $+38.4$ & $5955$ & $6791$ & $+14.0$ \\
O$_{5,6,7}$ & $1.586\times 10^{-3}$ & $1.818\times 10^{-3}$ & $+14.6$ & $9231$ & $10172$ & $+10.2$ \\
\hline
\end{tabular}
}
\caption{Validation of the additivity scheme using the  Eu(OH)$_3$ molecules.  For each Eu(OH)$_3$ structure, the $d_{00}$ and $\mathcal{W}_\mrm{NSM}$ values (a.u.) obtained directly from the calculations (denoted with subscripts ``dir'') are compared with the values obtained using the additivity scheme (denoted with subscripts ``add'') that sums the parameters of the 
three corresponding Eu(OH)$^{2+}$ complexes listed in Table~\ref{tab:Opos}. The fourth and the last columns show the percentage deviations $\Delta X=(X_{\mathrm{add}}/X_{\mathrm{dir}}-1)\times 100\%$.}
\label{tab:euoh3_add}
\end{table}

\end{document}